\documentclass[prc,aps,twocolumn,showpacs,superscriptaddress]{revtex4-1}

\usepackage{array}
\usepackage{amsmath}
\usepackage{graphicx}
\usepackage{amssymb}

\usepackage{color}
\usepackage[colorlinks,citecolor=blue]{hyperref}
\usepackage{subfigure}
\usepackage{bm}
\usepackage[colorinlistoftodos]{todonotes}
\usepackage{mathtools}

\usepackage{xcolor}
\newcommand{\orcid}[1]{\href{https://orcid.org/#1}{\textcolor[HTML]{A6CE39}{\aiOrcid}}}

\providecommand{\aap}{Astron. Astrophys.}
\providecommand{\apjl}{Astrophys. J.}
\providecommand{\apjs}{Astrophys. J. Suppl.}
\providecommand{\physrep}{Phys. Rep.}
\providecommand{\pasa}{Publ. Astron. Soc. Austr.}

\makeatletter

\newcommand*\@dblLabelI {}
\newcommand*\@dblLabelII {}
\newcommand*\@dblequationAux {}

\def\@dblequationAux #1,#2,%
    {\def\@dblLabelI{\label{#1}}\def\@dblLabelII{\label{#2}}}

\newcommand*{\doubleequation}[3][]{%
    \par\vskip\abovedisplayskip\noindent
    \if\relax\detokenize{#1}\relax
       \let\@dblLabelI\@empty
       \let\@dblLabelII\@empty
    \else 
       \@dblequationAux #1,%
    \fi
    \makebox[0.5\linewidth-3.5em]{%
     \hspace{\stretch2}%
     \makebox[0pt]{$\displaystyle #2$}%
     \hspace{\stretch1}%
    }%
    \makebox[0.5\linewidth+0.5em]{%
     \hspace{\stretch1}%
     \makebox[0pt]{$\displaystyle #3$}%
     \hspace{\stretch2}%
    }%
    \makebox[3em][r]{(%
  \refstepcounter{equation}\theequation\@dblLabelI, 
  \refstepcounter{equation}\theequation\@dblLabelII)}%
  \par\vskip\belowdisplayskip
}

\begin{document}

\title{Neutrino signal from proto-neutron star evolution: Effects of opacities from charged-current--neutrino interactions and inverse neutron decay}

\author{Tobias Fischer}\email{tobias.fischer@uwr.edu.pl}
\affiliation{Institute for Theoretical Physics, University of Wroc{\l}aw, 50-204 Wroc{\l}aw, Poland}

\author{Gang Guo}\email{gangg23@gmail.com}
\affiliation{GSI Helmholtzzentrum f\"ur Schwerioneneforschung, 64291 Darmstadt, Germany}
\affiliation{Institute of Physics, Academia Sinica, Taipei, 11529, Taiwan}

\author{Alan A. Dzhioev}
\affiliation{Bogoliubov Laboratory of Theoretical Physics, Joint Institute for Nuclear Research, 141980 Dubna, Moscow region, Russia}

\author{Gabriel Mart{\'i}nez-Pinedo}
\affiliation{GSI Helmholtzzentrum f\"ur Schwerioneneforschung, 64291 Darmstadt, Germany}
\affiliation{Technische Universit{\"a}t Darmstadt, 64289 Darmstadt, Germany}

\author{Meng-Ru Wu}
\affiliation{Institute of Physics, Academia Sinica, Taipei, 11529, Taiwan}
\affiliation{Institute of Astronomy and Astrophysics, Academia Sinica, Taipei, 10617, Taiwan}

\author{Andreas Lohs}
\affiliation{GSI Helmholtzzentrum f\"ur Schwerioneneforschung, 64291 Darmstadt, Germany}

\author{Yong-Zhong Qian}
\affiliation{School of Physics and Astronomy, University of Minnesota, Minneapolis, MN 55455, USA}

\begin{abstract}
We investigate the impact of charged-current neutrino processes on the formation and evolution of neutrino spectra during the deleptonization of proto-neutron stars. To this end we develop the full kinematics of these reaction rates consistent with the nuclear equation of state, including weak magnetism contributions. This allows us to systematically study the impact of inelastic contributions and weak magnetism on the $\nu_e$ and $\bar\nu_e$ luminosities and average energies. Furthermore, we explore the role of the inverse neutron decay, also known as the direct Urca process, on the emitted spectra of $\bar\nu_e$. This process is commonly considered in the cooling scenario of cold neutron stars but has so far been neglected in the evolution of hot proto-neutron stars. We find that the inverse neutron decay becomes the dominating opacity source for low-energy $\bar\nu_e$. Accurate three-flavor Boltzmann neutrino transport enables us to relate the magnitude of neutrino fluxes and spectra to details of the treatment of weak processes. This allows us to quantify the corresponding impact on the conditions relevant for the nucleosynthesis in the neutrino-driven wind, which is ejected from the proto-neutron star surface during the deleptonization phase. 
\end{abstract}

\received{28 April 2018} \revised{manuscript received 13 November 2019} \accepted{28 January 2020}

\maketitle

\section{Introduction}
\label{intro}
In this paper we study the impact of opacities from weak processes on the emission of neutrinos from newly-born proto-neutron stars (PNS). A PNS forms in the event of a core-collapse supernova, when the collapsing stellar core of a massive star bounces back at supra-saturation density. A hydrodynamics shock forms, which initially propagates quickly to large radii on the order of 100~km, where it stalls due to energy losses and turns into an accretion front. The revival of this stalled shock results in the supernova explosion and the subsequent ejection of the stellar mantle. The revival of the shock via neutrino heating is considered one of the standard explosion mechanisms~\cite{Bethe:1985ux}, besides the magneto-rotational~\cite{LeBlanc:1970kg} and acoustic mechanisms~\cite{Burrows:2005dv} as well as those driven by a high-density phase transition~\cite{Sagert:2008ka,Fischer:2011,Fischer:2018} (see also Refs.~\cite{Janka:2007,Janka:2012,Janka.Melson.Summa:2016} for recent reviews about core-collapse supernovae). Being initially hot and lepton rich, the PNS contains all gravitational binding energy gain of the collapsed stellar core. This energy is emitted mainly in the form of neutrinos on a timescale of $\simeq$10--30~s, once the supernova explosion proceeds.

Shortly after the onset of the supernova explosion, the PNS enters the deleptonization phase, which is determined by the diffusion of neutrinos of all flavors from high densities. These neutrinos decouple from matter at their spheres of last scattering, outside which they are free-streaming and become a potentially observable signal~\cite{Fischer:2012a,Wu:2015,Fischer:2016b}. The PNS deleptonization phase has long been studied \cite{Pons:1998mm,Page:2000}, with the commonly employed diffusion approximation for neutrino transport within the hydro-static approach. Only recently it has been explored in radiation-hydrodynamics simulations that employed accurate three-flavor Boltzmann neutrino transport~\cite{Fischer:2009af,Huedepohl:2010}. These studies followed the entire evolution consistently, i.e. starting from stellar core collapse and through core bounce, supernova explosion and PNS deleptonization. As the currently operating and future planned neutrino detectors are capable of detecting thousands of events from the next Galactic supernova~\cite{Dasgupta:2010,Serpico:2012}, it is of paramount interest to predict accurate neutrino luminosities and spectra. The primary focus is therefore on the development of models that include the leading weak processes and implement them in accord with the description of the nuclear medium~\cite{Reddy:1998,MartinezPinedo:2012,Roberts.Reddy.Shen:2012}. At the mean-field level, medium modifications for the charged-current reactions depend on the nuclear equation of state (EOS), in particular on the nuclear symmetry energy and its density dependence~\cite{Typel:2014}. The symmetry energy has been shown to affect the deleptonization timescale of the PNS~\cite{Roberts.Shen.ea:2012}.

High neutrino luminosities on the order of $10^{51}$~erg~s$^{-1}$ during the PNS cooling phase are responsible for the production of an outflow of matter known as the neutrino-driven wind. The latter has long been studied as a possible site for the nucleosynthesis of heavy elements~\cite{Woosley:1994ux,Takahashi:1994yz}. The proton-to-nucleon ratio, or equivalently the electron fraction $Y_e$, of the neutrino-driven wind depends sensitively on the spectral difference between $\bar\nu_e$ and $\nu_e$~\cite{Qian:1996xt}, requiring accurate predictions of these spectra. In this work, we determine how the spectra are affected by weak magnetism and nucleon recoil corrections. Therefore, we develop charged-current rate expressions within the relativistic approach taking into account the full kinematics and including contributions from weak magnetism~\cite{MartinezPinedo:2019}, being complementary to Ref.~\cite{Roberts.Reddy:2017}. This allows us to compare quantitatively with the elastic approximation as well as Ref.~\cite{Horowitz:2001xf} which approximates inelastic contributions and weak magnetism corrections. Furthermore, we include here the impact of the inverse neutron decay. Up to now, this opacity channel has never been considered during the evolution of PNSs. In Sec.~\ref{sec:ejecta}, we explore the impact of its inclusion together with a full treatment of charged-current processes on the composition of the ejecta, extending previous simulations by us~\cite{MartinezPinedo:2014}.

The paper is organized as follows. Section~\ref{sec:inverse-neutr-decay} presents basic expressions for the treatment of the charged-current neutrino opacity and in particular for neutron decay and its inverse reaction, derived from the full kinematics approach including weak magnetism contributions. In Sec.~\ref{sec:supernova-model} we review our supernova model {\tt AGILE-BOLTZTRAN} and in Sec.~\ref{sec:pns-evolution} we discuss simulations of the PNS deleptonization with particular focus on the impact of weak magnetism corrections and nucleon recoil contributions as well as the inclusion of the neutron decay channel, on the formation and evolution of the neutrino fluxes and spectra. The relevance for the nucleosynthesis of the neutrino-driven wind is discussed in Sec.~\ref{sec:ejecta}. A summary is given in Sec.~\ref{sec:summary}. 

\section{Charged-current processes and inverse neutron decay opacity}
\label{sec:inverse-neutr-decay}
Let us consider a generic charged-current process
\begin{equation}
\label{eq:1breaction}
\nu_1 + B_2 \longrightarrow l_3 + B_4 
\end{equation}
where $\nu_1$ denotes a neutrino (antineutrino), $B_2$ a neutron (proton), $l_3$ an electron (positron) and $B_4$ a proton (neutron). For the energies considered the interaction Lagrangian from the Weinberg-Salam theory~\cite{Glashow:1961,Weinberg:1967,Salam:1968} can be reduced to a current-current
interaction~\cite{Donnelly.Peccei:1979,Walecka:1975,Reddy:1998}:
\begin{equation}
\label{eq:lagrangian}
\mathcal{L} = \frac{G}{\sqrt{2}} l_\mu j^\mu 
\end{equation}
where $G=G_F V_{ud}$ for charged-current processes and $G=G_F$ for neutral-current processes, with $G_F$ the Fermi coupling constant and $V_{ud}$ the up-down entry of the Cabibbo-Kobayashi-Maskawa matrix~\cite{Tanabashi.Others:2018}. The lepton weak current is
\begin{equation}
\label{eq:leptoncurrent}
l_\mu = \bar{\psi}_3\gamma_\mu (1-\gamma_5)\psi_1
\end{equation}
and the hadronic current
\begin{equation}
\label{eq:hadroncurrent}
j^\mu = \bar{\psi}_4\left[\gamma^\mu\left(g_V-g_A\gamma_5\right) + \frac{i F_2}{2 M} \sigma^{\mu\nu} q^*_\nu\right] \psi_2 
\end{equation}
where $\psi_i$ are the Dirac spinors, and $g_V=1.0$, $g_A=1.273$, and $F_2=3.706$ are the coupling constants for the vector current, axial vector current, and weak magnetism, respectively. We do not consider the induced pseudoscalar term as its contribution is negligible to the processes considered here.

\subsection{Full kinematics}
At the mean-field or Hartree level considered here the nucleons fulfill the energy-momentum relationship: $E_{2,4}^* = E_{2,4} - U_{2,4} = \sqrt{|\bm{p}_{2,4}|^2+(m_{2,4}^{*})^2}$, with $U_{2,4}$ the mean-field potential and $m_{2,4}^{*}$ the effective mass. Introducing the four-momenta $p^*_{2,4} = (E^*_{2,4}, \bm{p}_{2,4})$, the four-momentum transfer becomes $q^* = p^*_{4} - p^*_{2}$. Notice that in the weak-magnetism term we use $q^*$ instead of $q$ as required by conservation of the weak vector current~\cite{Leinson.Perez:2001,Leinson:2002,Roberts.Reddy:2017}. Note also that in the following we will use natural units $\hbar=c=1$.

\begin{widetext}
The opacity or inverse mean-free path for process \eqref{eq:1breaction} is given by the integral expression
\begin{equation}
\label{eq:opacity_full}
\chi(E_1)= 2 \int\frac{d^3p_2}{(2\pi)^3}\int\frac{d^3p_3}{(2\pi)^3}\int\frac{d^3p_4}{(2\pi)^3} \frac{\langle\left\vert\mathcal{M}\right\vert^2\rangle}{16E_1E_2E_3E_4}\left(1-f_3\right)f_2\left(1-f_4\right)\left(2\pi\right)^4\delta^{(4)}(p_1+p_2-p_3-p_4)~,
\end{equation}
where the baryon ($B_2, B_4$) and the lepton ($l_3$) contributions enter via their corresponding thermal equilibrium Fermi distributions, $f_i=\left[\exp\{\beta(E-\mu_i)\}+1\right]^{-1}$, with the inverse temperature $\beta=1/T$. The spin-averaged and squared matrix element corresponding to the hadronic current~\eqref{eq:hadroncurrent} is given as
\begin{equation}
\label{eq:M}\langle\left\vert\mathcal{M}\right\vert^2\rangle =  \langle\left\vert\mathcal{M}_0\right\vert^2\rangle + \langle\left\vert\mathcal{M}\right\vert^2\rangle_{VF} \pm \langle\left\vert\mathcal{M}\right\vert^2\rangle_{FA} + \langle\left\vert\mathcal{M}\right\vert^2\rangle_{FF}~,
\end{equation}
where the terms associated with $\mathcal{M}_0$ correspond to the hadronic current without weak magnetism ($F_2=0$),
\begin{equation}
\label{eq:M0}\langle\left\vert\mathcal{M}_0\right\vert^2\rangle =  \langle\left\vert\mathcal{M}\right\vert^2\rangle_{VV} + \langle\left\vert\mathcal{M}\right\vert^2\rangle_{AA} \pm \langle\left\vert\mathcal{M}\right\vert^2\rangle_{VA}~,\\
\end{equation}
with
\begin{subequations}
\begin{eqnarray}
\label{eq:m}
\label{eq:M0VV}\langle\left\vert\mathcal{M}\right\vert^2\rangle_{VV} &=&  \left(4\,G\right)^2 g_V^2 \left[\left(p_1\cdot p_2^*\right)\left(p_3\cdot p_4^*\right)+\left(p_1\cdot p_4^*\right)\left(p_3\cdot p_2^*\right)-m^*_2m^*_4\left(p_1\cdot p_3\right)\right]~,\\
\label{eq:M0AA}\langle\left\vert\mathcal{M}\right\vert^2\rangle_{AA} &=& \left(4\,G\right)^2 g_A^2 \left[\left(p_1\cdot p_2^*\right)\left(p_3\cdot p_4^*\right)+\left(p_1\cdot p_4^*\right)\left(p_3\cdot p_2^*\right)+m^*_2m^*_4\left(p_1\cdot p_3\right)\right]~, \\
\label{eq:M0VA}\langle\left\vert\mathcal{M}\right\vert^2\rangle_{VA} &=&  2\left(4\,G\right)^2 g_V g_A \left[\left(p_1\cdot p_2^*\right)\left(p_3\cdot p_4^*\right)-\left(p_1\cdot p_4^*\right)\left(p_3\cdot p_2^*\right)\right]~.
\end{eqnarray}
\end{subequations}
The remaining terms in Eq.~(\ref{eq:M}) are the contributions due to weak magnetism,
\begin{subequations}
\begin{eqnarray}
\label{eq:m}
  \langle\left\vert\mathcal{M}\right\vert^2\rangle_{VF} =
  (4\,G)^2 g_V \frac{F_2}{2M}\!&\bigl\{&\!\left[\left(p_1\cdot p_2^*\right)m_4^*-\left(p_1\cdot p_4^*\right)m_2^*\right]\left(p_3\cdot q^*\right)\nonumber\\
      & &+ \left[\left(p_3\cdot p_2^*\right)m_4^* - \left(p_3\cdot p_4^*\right)m_2^*\right]\left(p_1\cdot q^*\right) \nonumber\\
& & + \left[\left(q^*\cdot p_2^*\right)m_4^* - \left(q^*\cdot p_4^*\right)m_2^*\right]\left(p_1\cdot p_3\right)
\bigr\},\\
\langle\left\vert\mathcal{M}\right\vert^2\rangle_{FA} = 2\left(4\,G\right)^2 g_A \frac{F_2}{2M}\! &\bigl\{&\! \left[\left(p_1\cdot p_2^*\right)m_4^*+\left(p_1\cdot p_4^*\right)m_2^*\right]\left(p_3\cdot q^*\right) \nonumber\\
&& - \left. \left[\left(p_3\cdot p_2^*\right)m_4^* + \left(p_3\cdot p_4^*\right)m_2^*\right]\left(p_1\cdot q^*\right)
\right\}~,\\
\langle\left\vert\mathcal{M}\right\vert^2\rangle_{FF} = \frac{1}{2}\left(4\,G\right)^2 \left(\frac{F_2}{2M}\right)^2\!& \bigl\{&\! 2\left[\left(p_3\cdot p_2^*\right)\left(p_4^*\cdot q^*\right)+\left(p_3\cdot p_4^*\right)\left(p_2^*\cdot q^*\right)\right]\left(p_1\cdot q^*\right) \nonumber\\
& &+ 2\left[\left(p_1\cdot p_2^*\right)\left(p_4^*\cdot q^*\right)+\left(p_1\cdot p_4^*\right)\left(p_2^*\cdot q^*\right)\right]\left(p_3\cdot q^*\right) \nonumber\\
& &- 2\left(p_1\cdot q^*\right)\left(p_3\cdot q^*\right)\left(p_2^*\cdot p_4^*\right) \nonumber\\
& &+ q^{*2} \left[\left(p_1\cdot p_3\right)\left(p_2^*\cdot p_4^*\right)-2\left(p_1\cdot p_2^*\right)\left(p_3\cdot p_4^*\right)-2\left(p_1\cdot p_4^*\right)\left(p_3\cdot p_2^*\right)\right] \nonumber\\
& &- m_2^*m_4^*\left[\left(p_1\cdot p_3\right)q^{*2}+2\left(p_1\cdot q^*\right)\left(p_3\cdot q^*\right)\right]
\bigl\}.
\end{eqnarray}
\end{subequations}
The `$+$' (`$-$') sign of the $VA$ and $FA$ terms in Eqs.~\eqref{eq:M} and \eqref{eq:M0} corresponds to the neutrino (anti-neutrino) reaction. Applying the energy-momentum delta function and integrating over all angles, expression~\eqref{eq:opacity_full} can be reduced algebraically to a 2-dimensional integral as follows (details are given in Appendix~\ref{sec:X}),
\begin{equation}
\label{eq:opacity_final}
\chi(E_1) = \frac{G^2}{4\pi^3E_1^2} \int_{E_{3^-}}^{E_{3^+}}dE_3\left(1-f_3\right)\int_{E_{2^-}}^{E_{2^+}}dE_2 f_2\left(1-f_4\right)~\Phi~.
\end{equation}
The explicit form of $\Phi$, as well as the integration limits $E_{2^\pm}$ and $E_{3^\pm}$, can be found in Appendix~\ref{sec:X}.

\subsection{Inverse neutron decay}
For cold neutron star conditions, neutron decay, $n \rightarrow p + e^- + \bar\nu_e$, being part of the direct Urca processes, is suppressed whenever the proton faction $Y_p < x_c$ with $x_c = 1/9$ if muons are neglected and $x_c\simeq 0.148$ if they are considered~\cite{Lattimer.Pethick.ea:1991,Yakovlev.Pethick:2004}. For PNS evolution, due to the higher temperature on the order of tens of MeV and higher $Y_e$ values, neutron decay may not be ignored as discussed below.

We can generalize the rate expression~\eqref{eq:opacity_final} to describe the $\bar\nu_e$--opacity for the inverse neutron decay, $\bar{\nu}_e + e^- + p \rightarrow n$, as follows:
\begin{equation}
\label{eq:opa_dec}
\chi^{\text{dec}}(E_{\bar\nu_e}) = \frac{G^2}{4\pi^3 E_{\bar\nu_e}^2} \int_{E_{3^-}}^{E_{3^+}}dE_ef_e \int_{E_{2^-}}^{E_{2^+}}dE_p f_p\left(1-f_n\right)~\Phi^{\text{dec}}~, 
\end{equation}
where $\Phi^{\text{dec}}$ is obtained from $\Phi$ for $\bar\nu_e$
capture on protons with the replacement $E_3 \to -E_3$ (see
Appendices~\ref{sec:X} and \ref{app:B}).

\end{widetext}

\subsection{Elastic approximation}
It is convenient to introduce the elastic approximation, which has been commonly employed in supernova studies. This approximation is obtained by considering non-relativistic nucleons, the dominant $VV$ and $AA$ terms in the matrix element neglecting dependence on the lepton angle, and assuming zero momentum transfer. Under these assumptions the opacity reduces to \cite{Bruenn:1985en,Reddy:1998}:
\begin{equation}
\label{eq:opelastic}
\chi_0(E_1) \approx \frac{G^2}{\pi} \left(g_V^2 + 3 g_A^2\right)\, p_3\,E_3\,\left[1-f_3(E_3)\right]\,\eta_{24}~,
\end{equation}
with
\begin{equation}
\label{eq:eta}
\eta_{24} = \frac{n_2-n_4}{1-\exp[\beta(\varphi_4-\varphi_2)]}~,
\end{equation}
where $n_i$ denote the number density of particle species $i$ and $\varphi_i$ are the free Fermi gas chemical potential, related to the nuclear EOS chemical potentials $\mu_i$ via $\varphi_i = \mu_i - m_i^* - U_i$.

\begin{figure}[htp]
\centering
\includegraphics[width=1.0\columnwidth]{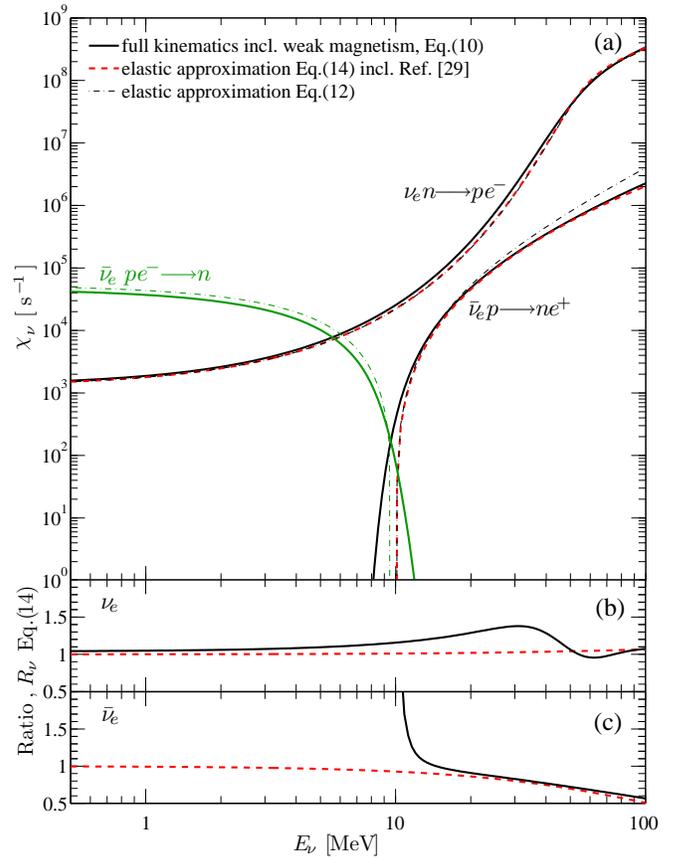}
\caption{(Color online) {\it Top panel (a)}: Opacity for charged-current reactions at selected conditions (temperature $T=7$~MeV, density $\rho=2\times 10^{13}$~g~cm$^{-3}$, electron fraction $Y_e=0.05$) corresponding to a mean-filed potential difference of $U_n-U_p=8.3$~MeV. Shown are the processes 1 and 2 of Table~\ref{tab:nu-reactions} for the full kinematics approach including weak magnetism Eq.~\eqref{eq:opacity_final} (thick solid black lines), in comparison with the elastic approximation  Eq.~\eqref{eq:opelastic} (thin black dash-dotted lines) and including the approximate treatment of inelasticity and weak magnetism of Ref.~\cite{Horowitz:2001xf} Eqs.~\eqref{eq:R} and \eqref{eq:Rbar} (thick dashed red lines). The inverse neutron decay opacity for $\bar\nu_e$ is shown via the green lines in the same setup, full kinematics Eq.~\eqref{eq:opa_dec} (thick solid line) vs. elastic approximation Eq.~\eqref{eq:invndecayelas} (thin dash dotted line). {\it Bottom panels (b) and (c)}: approximate correction factors for inelastic contributions and weak magnetism \eqref{eq:R} for $\nu_e$ and \eqref{eq:Rbar} for $\bar\nu_e$ of Ref.~\cite{Horowitz:2001xf} (red dashed lines), in comparison to the ratios of Eqs.~\eqref{eq:opacity_final} and \eqref{eq:opelastic} (solid black lines).}
\label{fig1}
\end{figure}

Within this approximation, inelastic contributions to the opacity and corrections due to weak magnetism are commonly taken into account based on the description provided in Ref.~\cite{Horowitz:2001xf}, where these effects are approximated by neutrino energy dependent factors for $\nu_e$ and $\bar\nu_e$,
\begin{subequations}
\begin{eqnarray}
\label{eq:R} && \chi(E_{\nu_e}) = R_{\nu_e}(E_{\nu_e})~\chi_0(E_{\nu_e})~,\\
\label{eq:Rbar} && \chi(E_{\bar\nu_e}) = R_{\bar\nu_e}(E_{\bar\nu_e})~\chi_0(E_{\bar\nu_e})~.
\end{eqnarray}
\end{subequations}
The factors $R_{\nu_e}(E_{\nu_e})$ and $R_{\bar\nu_e}(E_{\bar\nu_e})$ are shown in the bottom panels of Fig.~\ref{fig1}.

Let us first consider $\nu_e$ absorption on neutrons. At low neutrino energies such that $E_e < \mu_e$, the opacity is suppressed by final state blocking and behaves as
\begin{equation}
\label{eq:5}
\chi_0(E_{\nu_e}) \approx (E_{\nu_e} +\Delta m^*_{np} +\Delta U_{np})^2 \, e^{\beta(E_{\nu_e} +\Delta m^*_{np} +\Delta U_{np}-\mu_e)}~,
\end{equation}
with $\Delta m^*_{np}=m_n^*-m_p^*$ and $\Delta U_{np}=U_n-U_p$. From expression~\eqref{eq:5} it becomes clear that the opacity increases exponentially with $\Delta U_{np}$, i.e. with increasing symmetry energy. 

For $\bar{\nu}_e$, there is no final state blocking for the positron. The main effect of the mean-field correction $\Delta U_{np}$ is to suppress the opacity for antineutrino energies $E_{\bar{\nu}_e} < m_e + \Delta m^*_{np}+\Delta U_{np}$, or $E_{\bar{\nu}_e} \lesssim 10$~MeV for typical conditions around the $\bar{\nu}_e$ sphere of last scattering. Including inelastic contributions to the response function will result in a non-zero opacity at energies lower than the above threshold, as illustrated in Fig.~\ref{fig1} comparing the solid black line (full kinematics) and dash-dotted line (elastic approximation). However, such opacity is strongly suppressed as also seen in Fig.~\ref{fig1}. This behavior has been related to the frustrated kinematics due to energy-momentum conservation restrictions on single-particle interactions in processes similar to~\eqref{eq:1breaction} (cf. \cite{Roberts.Reddy.Shen:2012}). The authors of Ref.~\cite{Roberts.Reddy.Shen:2012} explored the role of nuclear many-body correlations at the level of the random-phase approximation and reactions with two nucleons in the initial and final state (e.g. $\nu_e + n + n \rightarrow n + p + e^-$ and $\bar{\nu}_e + p + n \rightarrow n + n + e^+$), and found small corrections. These charged-current processes are generalizations of the modified-Urca reactions that are the main source of neutrino production in cold neutron stars~\cite{Friman.Maxwell:1979,Yakovlev.Kaminker.ea:2001}.

The opacity for inverse neutron decay can also be obtained within the elastic approximation:
\begin{equation}
\label{eq:invndecayelas}
\chi^{\text{dec}}_0(E_{\bar{\nu}_e}) \approx \frac{G^2}{\pi}\left(g_V^2 + 3 g_A^2\right) p_e E_e \, f(E_e) \, \eta_{pn}~,
\end{equation}
with $E_e = \Delta m^*_{np} + \Delta U_{np} - E_{\bar{\nu}_e}$. 

Figure~\ref{fig1} shows that for low neutrino energies the $\bar\nu_e$--opacity due to inverse neutron decay and the $\nu_e$--opacity for absorption on neutrons are of similar magnitude. This is a general result that is independent from the treatment of the nuclear medium as demonstrated in appendix~\ref{app:B}).

The equations of state commonly used in core-collapse supernova simulations treat neutrons and protons as non-interacting quasi-particles that move in a mean-field single particle potential, $U_{2,4}$, that depends on temperature, density and $Y_e$. For the neutron-rich conditions around the neutrinospheres the mean-field potentials for protons and neutrons can be very different, on the order of tens of MeV. The difference $\Delta U=U_2-U_4$ is directly related to the nuclear symmetry energy~\cite{Fischer:2014,Hempel:2015a}, which has a strong density dependence. Note that for practical purposes, we integrate Eq.~\eqref{eq:opacity_final} numerically using 32 grid point Gauss quadratures for each dimension.

\section{Supernova model}
\label{sec:supernova-model}
Our core-collapse supernova model, {\tt AGILE-BOLTZTRAN}, is based on spherically symmetric and general relativistic neutrino radiation hydrodynamics with angle- and energy-dependent three flavor Boltzmann neutrino transport~\cite{Liebendoerfer:2001a,Liebendoerfer:2001b,Liebendoerfer:2002}. The implicit method for solving the hydrodynamics equations and the Boltzmann transport equation on an adaptive Lagrangian mesh has been compared with other methods, e.g., with the multi-group flux limited diffusion approximation~\cite{Liebendoerfer:2004} and the variable Eddington factor technique \cite{Liebendoerfer:2005a}.

Here we employ the nuclear EOS from Ref.~\cite{Hempel:2009mc}. It is based on the relativistic mean-field (RMF) framework for homogeneous nuclear matter with the RMF parametrization DD2~\cite{Typel:2005} henceforth denoted as HS(DD2). Moreover, nuclei are treated within the modified nuclear statistical equilibrium approach for several 1000 nuclear species based on tabulated and partly calculated masses. It is part of the comprehensive multi-purpose EOS catalogue {\tt COMPOSE}~\cite{Typel:2013rza}. In addition, lepton and photon contributions are added based on the EOS from Ref.~\cite{Timmes:1999}.

\begin{table}[t!]
\centering
\caption{Neutrino reactions considered, including references.}
\begin{tabular}{ccc}
\hline
\hline
& Weak process & Reference \\
\hline
1 & $e^- + p \rightleftarrows n + \nu_e$ & \cite{Reddy:1998,Horowitz:2001xf} \\ 
2 & $e^+ + n \rightleftarrows p + \bar\nu_e$ & \cite{Reddy:1998,Horowitz:2001xf} \\
3 & $n \rightleftarrows p + e^- + \bar\nu_e$ & this work \\
4 & $\nu_e + (A,Z-1) \rightleftarrows (A,Z) + e^-$ & \cite{Juodagalvis:2010} \\
5 & $\nu + N \rightleftarrows \nu' + N$ & \cite{Bruenn:1985en,Mezzacappa:1993gm,Horowitz:2001xf} \\
6 & $\nu + (A,Z) \rightleftarrows \nu' + (A,Z)$ & \cite{Bruenn:1985en,Mezzacappa:1993gm} \\
7 & $\nu + e^\pm \rightleftarrows \nu' + e^\pm$ & \cite{Bruenn:1985en}, \cite{Mezzacappa:1993gx} \\
8 & $e^- + e^+ \rightleftarrows  \nu + \bar{\nu}$ & \cite{Bruenn:1985en} \\
9 & $N + N \rightleftarrows  \nu + \bar{\nu} + N + N $ & \cite{Hannestad:1997gc} \\
10 & $\nu_e + \bar\nu_e \rightleftarrows  \nu_{\mu/\tau} + \bar\nu_{\mu/\tau}$ & \cite{Buras:2002wt,Fischer:2009} \\
11 & $\nu + \bar\nu + (A,Z) \rightleftarrows (A,Z)^*$ & \cite{Fuller:1991,Fischer:2013} \\
\hline
\end{tabular}
\\
$\nu=\{\nu_e,\bar{\nu}_e,\nu_{\mu/\tau},\bar{\nu}_{\mu/\tau}\}$ and $N=\{n,p\}$
\label{tab:nu-reactions}
\end{table}

The set of weak reactions considered is listed in Table~\ref{tab:nu-reactions}. For both charged-current absorption (reactions 1 and 2 in Table~\ref{tab:nu-reactions}) and neutral current scattering (reaction 5 in Table~\ref{tab:nu-reactions}) on nucleons, previously we employed the elastic approximation of Ref.~\cite{Bruenn:1985en}. For the consistent treatment of nuclear EOS and charged-current absorption processes 1 and 2 of Table~\ref{tab:nu-reactions}, medium modifications are take into account at the mean-field level following Ref.~\cite{Reddy:1998} based on the nuclear EOS HS(DD2). They determine spectral differences between $\nu_e$ and $\bar\nu_e$~\cite{MartinezPinedo:2012,Roberts.Reddy.Shen:2012,Horowitz:2012}. In this work we implement the newly developed charged-current rates with the full kinematics, expression~\eqref{eq:opacity_final}, into the charged-current interaction module of {\tt AGILE-BOLTZTRAN}. This extends the previously employed elastic approach, with the approximate treatment of recoil and weak magnetism corrections of Ref.~\cite{Horowitz:2001xf}. In order to accurately capture the low- and intermediate-energy behavior of the charged-current weak rates the neutrino energy resolution of {\tt BOLTZTRAN} is enhanced to 24 bins extended down to 0.5~MeV. For the neutral current neutrino nucleon scattering processes (5) in Table~\ref{tab:nu-reactions}, inelastic contributions and weak magnetism corrections are taken into account following Ref.~\cite{Horowitz:2001xf} for the elastic rates. The latter corrections were determined for conditions relevant to neutrino-driven wind ejecta where the initial nucleon can be assumed at rest and final state blocking can be neglected. It remains to be demonstrated to what extent they remain valid for the high density/temperature conditions at which neutrinos decouple. While recoil corrections are known to reduce similarly the $\nu_e$ and $\bar\nu_e$--opacity, weak magnetism corrections increase slightly the opacity for neutrinos and strongly reduce that of antineutrinos. Both effects have been commonly included in core-collapse supernova simulations (cf. Refs.~\cite{Buras:2005rp,Liebendoerfer:2005a}). They were also included in simulations of the PNS deleptonization~\cite{Huedepohl:2010}. Note that contributions due to strange quark contents are not taken into account~\cite{Melson:2015}.

Figure~\ref{fig1} shows the opacities from the inverse neutron decay (green lines) and the charged-current absorption processes 1 and 2 in Table~\ref{tab:nu-reactions} (black/red lines), comparing the full kinematics approach with weak magnetism contributions, Eq.~\eqref{eq:opacity_final} for processes 1 and 2 in Table~\ref{tab:nu-reactions} and Eq.~\eqref{eq:opa_dec} for the inverse neutron decay (thick solid lines), elastic approximation including recoil and weak magnetism, Eqs.~\eqref{eq:R} and \eqref{eq:Rbar}~\cite{Horowitz:2001xf} (dashed lines), and the elastic rates without recoil and weak magnetism corrections, Eq.~\eqref{eq:opelastic} (dash-dotted lines). The conditions, $T=7$~MeV, $\rho=2\times 10^{13}$~g~cm$^{-3}$, and $Y_e=0.1$, correspond to the neutrinosphere during the deleptonization phase, where electron neutrinos have an equilibrium chemical potential of $\mu^{\text{eq}}_{\nu_e}=25$~MeV and where $\Delta U = 8.3$~MeV. Note that at these conditions, besides neutrons ($X_n=0.75$, $\mu_n-m_n=-1.6$~MeV, $U_n=35.34$~MeV, $m_n^*=899.3$~MeV) and protons ($X_p=0.02$, $\mu_p-m_p=-37.8$~MeV, $U_p=27.04$~MeV, $m_p^*=898.0$~MeV), there is still a non-negligible abundance of light nuclear clusters, e.g., deuteron ($X_{^2{\rm H}}=0.027$) and triton ($X_{^3{\rm H}}=0.1$), as well as heavy nuclei ($X_{(A,Z)}=0.08$) with average atomic mass $A\simeq75$ and charge $Z\simeq22$, based on the HS(DD2) EOS employed here (see also Ref.~\cite{Fischer:2017}). In Fig.~\ref{fig1} we can identify the suppression of the $\bar\nu_e$-opacity for reaction 2 in Table~\ref{tab:nu-reactions} at low neutrino energies discussed in the previous section. Current supernova models, which neglect the neutron decay, produce low energy $\bar{\nu}_e$ by $N$--$N$ bremsstrahlung and down-scattering of high energy neutrinos by nucleons and electrons. The former process is relatively inefficient due to final state blocking by trapped$\nu_e$. Figure~\ref{fig1} shows that the neutron decay can in fact be the dominating production channel for low energy $\bar{\nu}_e$. Moreover, Fig.~\ref{fig1} (bottom panels) highlights the importance of the correct kinematics which cannot be captured by the energy-dependent approximate factors of Ref.~\cite{Horowitz:2001xf}. Note in particular the underestimation of the absorption rate of $\bar\nu_e$ in the intermediate-energy region where thermal excitations of the nucleons give rise to a finite opacity, unlike for the elastic rate which is identical to zero for $E_{\bar\nu_e}<\triangle U_{np}+\triangle m_{np}^*$. This is also the reason for the sudden rise of the inelastic-to-elastic rate ratio around $E_\nu\simeq 10-15$~MeV in Fig.~\ref{fig1} (bottom panel), below which the comparison between elastic and inelastic rates breaks down.

\section{Protoneutron star deleptonization}
\label{sec:pns-evolution}

\begin{figure}[ht!]
\includegraphics[width=1.0\columnwidth]{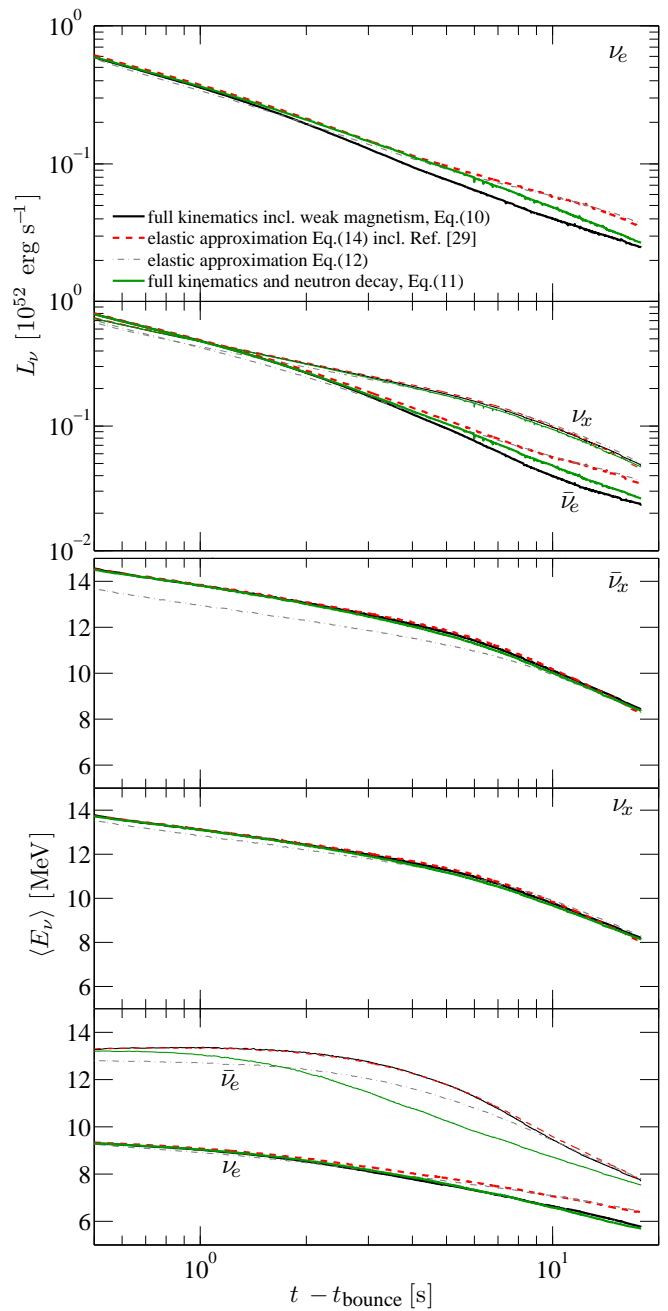}
\caption{(Color online) Post-bounce evolution of the neutrino energy luminosities $L_\nu$ and average energies $\langle E_\nu \rangle$ (sampled in the co-moving frame of reference at a radius of 1000~km), comparing four treatments of the charged-current rates: full kinematics (black lines), elastic approximation including inelastic corrections and weak magnetism contributions (red lines)~\cite{Horowitz:2001xf}, elastic approximation without inelastic corrections and weak magnetism contributions (grey lines) and full kinematics including the neutron decay channel (green lines). Note that $L_{\bar\nu_x}$ is not shown here because $L_{\bar\nu_x}$ and $L_{\nu_x}$ are nearly indistinguishable on the scale used.}
\label{fig:lumin}
\end{figure}
\begin{figure}[ht!]
\includegraphics[width=\columnwidth]{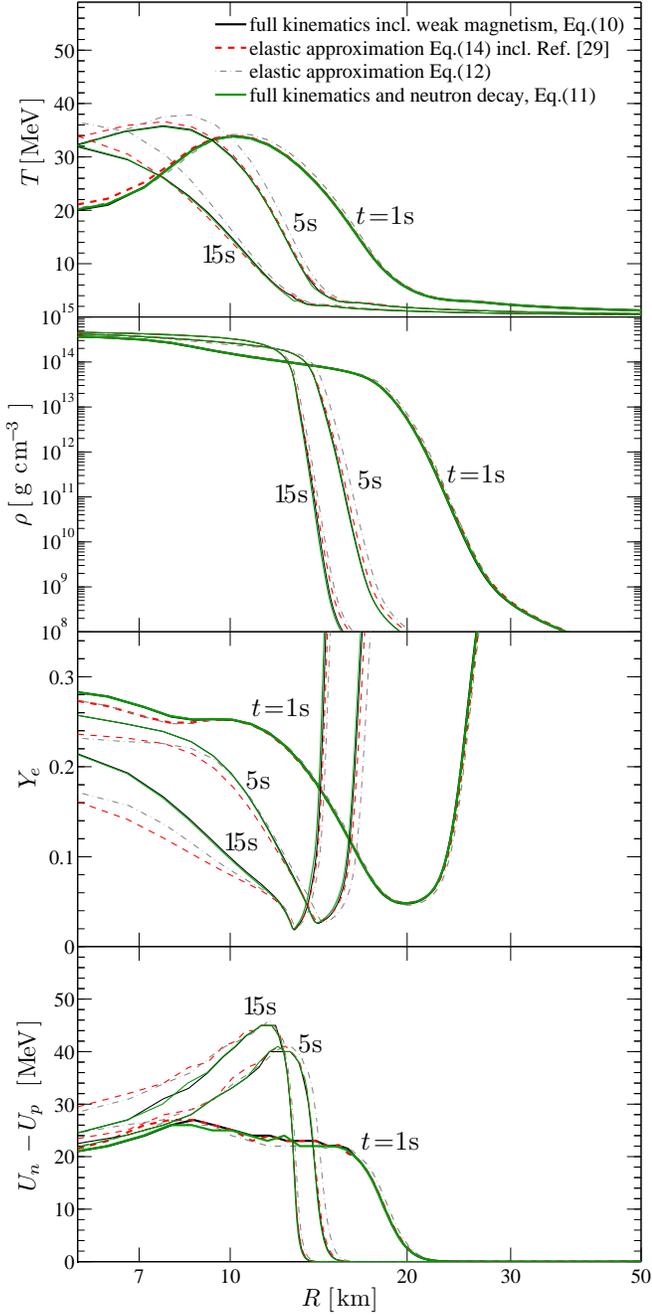}
\caption{(Color online) Evolution of the PNS structure at two selected post-bounce times, in terms of temperature $T$, density $\rho$, $Y_e$ and mean-field potential difference $U_n-U_p$, comparing four treatments of the charged-current rates: full kinematics (black lines), elastic approximation including inelastic corrections and weak magnetism contributions (red lines)~\cite{Horowitz:2001xf}, elastic approximation without inelastic corrections and weak magnetism contributions (grey lines) and full kinematics including the neutron decay channel (green lines).}
\label{fig:pns}
\end{figure}

In this section results from spherically symmetric supernova simulations are discussed. The simulations are launched from the 18~M$_\odot$ pre--collapse progenitor of Ref.~\cite{Woosley:2002zz}. It was evolved consistently through all phases prior to the supernova explosion onset, and has been subject to numerous supernova studies with focus on the stellar core collapse and bounce/post-bounce dynamics~\cite{Fischer:2013,Fischer:2016c,Fischer:2017} as well as simulations of the PNS deleptonization~\cite{Fischer:2016a,Fischer:2016b}. Here we focus on the latter in order to study the role of charged-current weak processes on the neutrino fluxes and spectra as well as their evolution after the supernova explosion onset. Since self-consistent neutrino-driven supernova explosions cannot be obtained in spherical symmetry, the explosion is triggered by artificially enhancing the neutrino heating via reactions~1 and 2 in Table~\ref{tab:nu-reactions}~\cite{Fischer:2009af}. After a short period of shock stalling, this results in the slow but continuous expansion of the bounce shock to increasingly larger radii, with the launch of the explosion onset at around $t=0.35$~s post bounce. The explosion shock reaches radii around 1000~km at about $t=0.5$~s post bounce. Once the explosion proceeds we switch back to the standard rates. This method has been employed previously~\cite{MartinezPinedo:2012} and compares qualitatively well with other artificially neutrino-driven explosion methods~\cite{Ugliano:2012,Perego:2015}. 

\begin{figure}[t!]
\includegraphics[width=0.975\columnwidth]{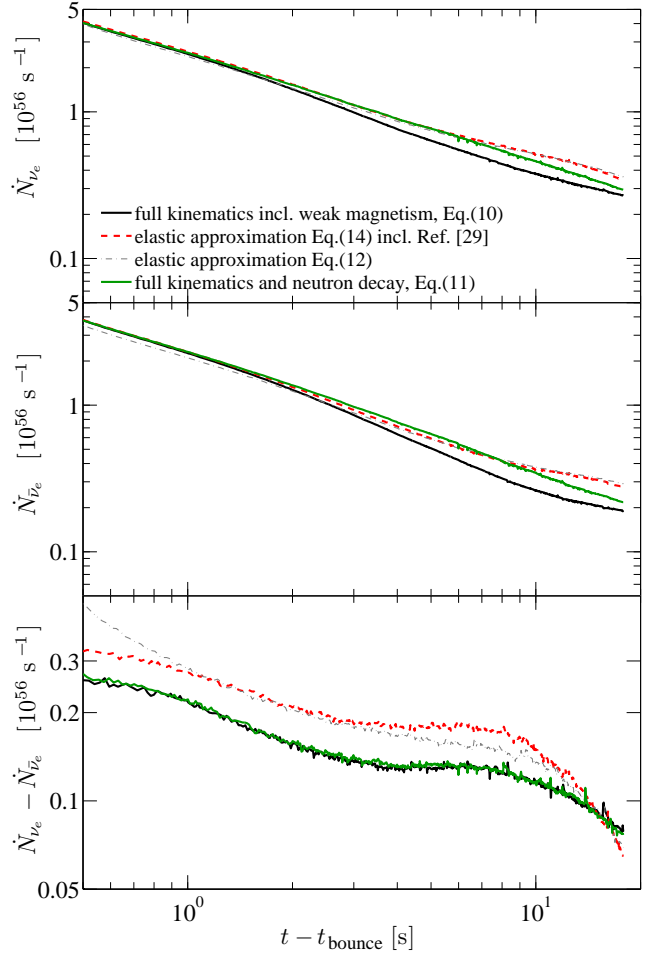}
\caption{(Color online) Post-bounce evolution of the $\nu_e$ and $\bar\nu_e$ number luminosity and deleptonization rate, $\dot N_{\nu_e} - \dot N_{\bar\nu_e}$, comparing four treatments of the charged-current rates: full kinematics (black lines), elastic approximation including inelastic corrections and weak magnetism contributions (red lines)~\cite{Horowitz:2001xf}, elastic approximation without inelastic corrections and weak magnetism contributions (grey lines) and full kinematics including the neutron decay channel (green lines).}
\label{fig:dnudt}
\end{figure}
\begin{figure*}[htb]
\includegraphics[width=2.0\columnwidth]{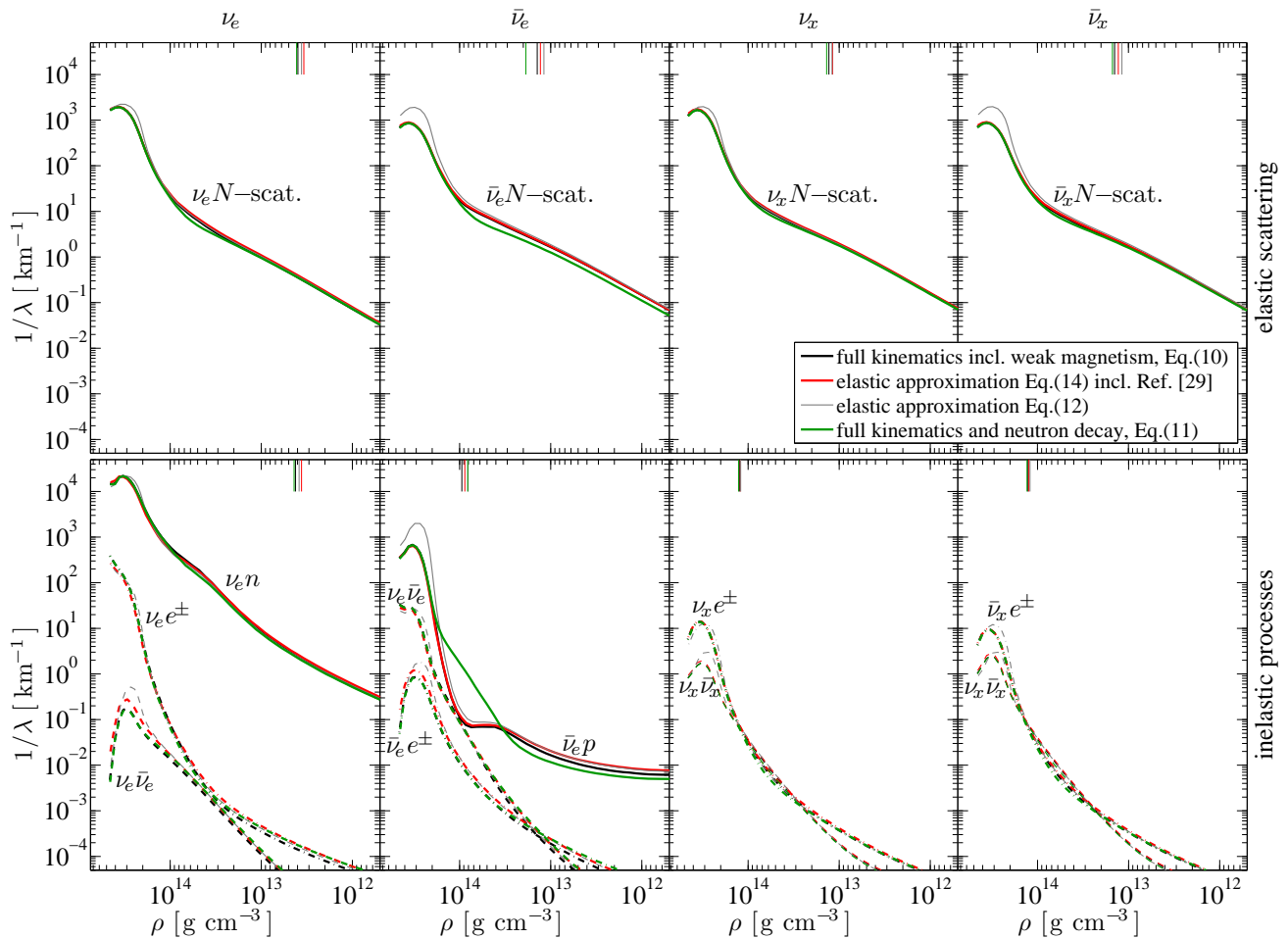}
\caption{(Color online) Density ($\rho$) dependence of the inverse neutrino mean-free path $1/\lambda$ for all flavors ($x=\mu/\tau$) and selected channels from  the simulations of the PNS deleptonization, comparing four treatments of the charged-current rates: full kinematics (black lines), elastic approximation including inelastic corrections and weak magnetism contributions (red lines)~\cite{Horowitz:2001xf}, elastic approximation without inelastic corrections and weak magnetism contributions (grey lines) and full kinematics including the neutron decay channel (green lines). The conditions correspond to the PNS deleptonization at a post-bounce time of 5~s (see also Fig.~\ref{fig:pns}). The opacity is separated into elastic scattering on nucleons (top) and inelastic processes (bottom). The channel $\nu_e\bar\nu_e$ ($\nu_x\bar\nu_x$) includes all of the pair processes 8--10 in Table~\ref{tab:nu-reactions} and the green line for the channel $\bar\nu_e p$ contains the inverse neutron decay. Vertical lines mark the position of the neutrinosphere of last elastic (top) and inelastic (bottom) reaction.}
\label{fig:mfp_a}
\end{figure*}
\begin{figure*}[htb]
\includegraphics[width=2.0\columnwidth]{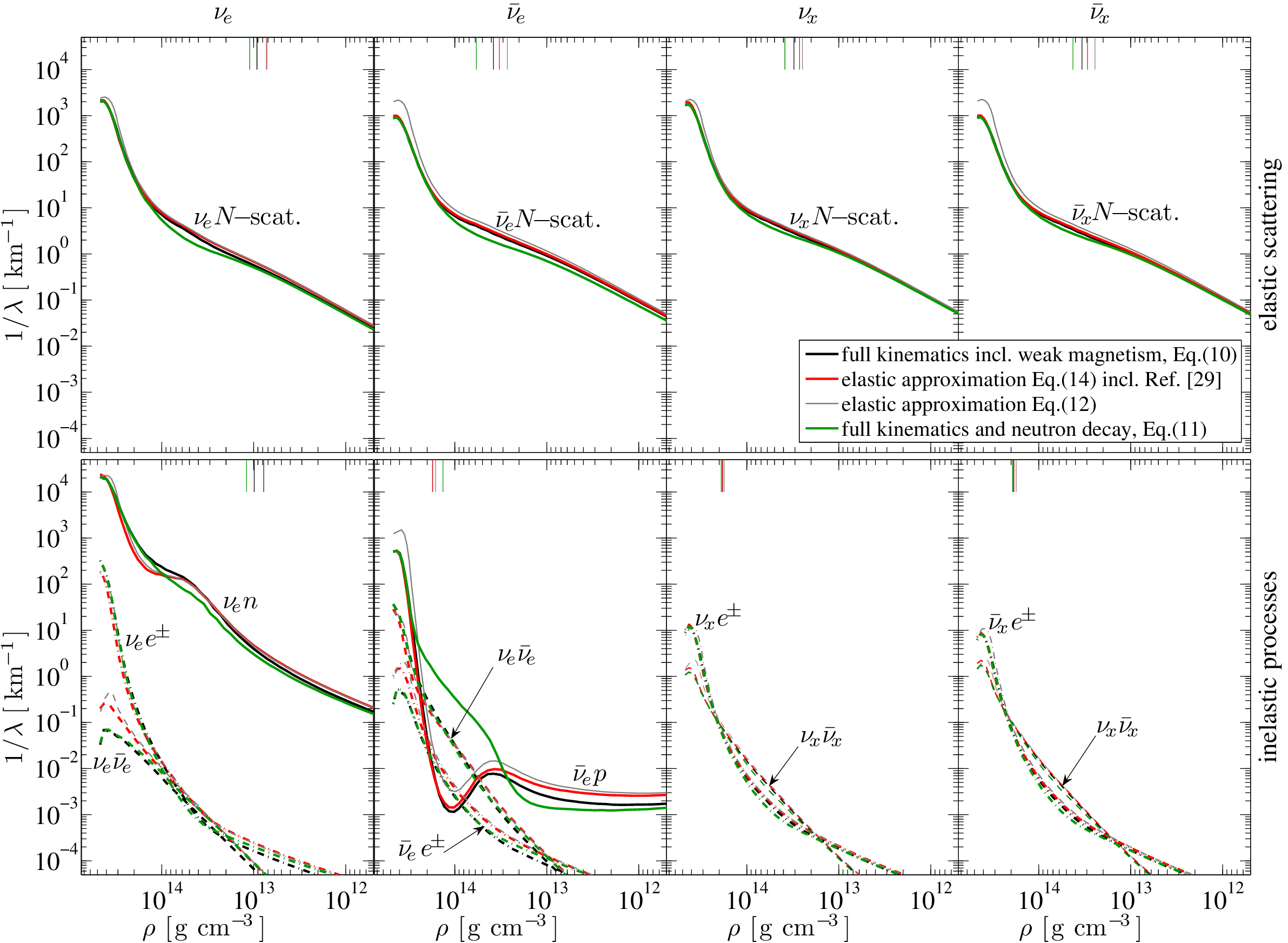}
\caption{(Color online) Same as Fig.~\ref{fig:mfp_a}, but for a post-bounce time of 10~s.}
\label{fig:mfp_b}
\end{figure*}

In order to quantify the impact of the different treatments of charged-current weak processes, we show in Fig.~\ref{fig:lumin} the post-bounce evolution of the neutrino luminosities (top panels) and average energies (bottom panels) for all flavors. We first compare the elastic approach (grey lines) and the approximate treatment of inelastic contributions and weak magnetism \cite{Horowitz:2001xf} (red lines), which show marginal differences in the luminosities of $\bar\nu_e$, $\nu_x$, and $\bar\nu_x$, as well as their average energies. In particular, employing the elastic treatment underestimates these quantities.  This is because the inclusion of weak magnetism and recoil suppresses the opacity of the charged-current interaction of $\bar\nu_e$ and neutral current interaction of all flavors~\cite{Horowitz:2001xf}. Note that as the effect due to the weak magnetism and recoil in Ref.~\cite{Horowitz:2001xf} is larger for higher neutrino energies, the differences discussed above decrease at later times when the average neutrino energies decrease.

In comparing these results now with those based on the full kinematics (black lines), Fig.~\ref{fig:lumin} shows that the $\nu_e$ and $\bar\nu_e$ luminosities and average energies are lower with the full kinematics treatment. This is due to the larger charged-current opacities (see Fig.~\ref{fig1}) at low and intermediate energies.  Nevertheless, the elastic approach with the approximate inclusion of weak magnetism and recoil provides a relatively good overall description of the gross evolution of the early PNS deleptonization. The main aspects, i.e. enhancement of the $\bar\nu$ fluxes and average energies are captured qualitatively, partly also because we employ the same elastic rates with the approximate inclusion of recoil and weak magnetism of Ref.~\cite{Horowitz:2001xf} for the neutrino-nucleon scattering rates \cite{Bruenn:1985en,Mezzacappa:1993gx} for all the models discussed here.

Because all four treatments of charged-current weak rates feature a similar evolution during the early PNS deleptonization phase, there are only marginal differences in the PNS restmass density and temperature structure at $t=1$~s as illustrated in Fig.~\ref{fig:pns}. However, the fully inelastic treatment results in a slower net deleptonization rate, ${\dot N}_{\nu_e}-{\dot N}_{\bar\nu_e}$, as shown in Fig.~\ref{fig:dnudt}  (black lines), already during the very early deleptonization phase on the order of few 100~ms. Note that prior to the supernova explosion onset the central evolution of the PNSs where weak equilibrium is established is identical for all simulations independent of the treatment of the charged-current weak rates. The situation changes shortly after the explosion onset, when the PNSs enter the deleptonization phase. The generally slower deleptonization rate for the full kinematics treatment, in comparison to the elastic rates (see Fig.~\ref{fig:dnudt}), features a higher $Y_e$ at high density domain of the PNS interior, which is notable already at about 1~s post bounce (see Fig.~\ref{fig:pns}). This difference increases during the proceeding PNS deleptonization. Note that the differences in the evolution of ${\dot N}_{\nu_e}$ and ${\dot N}_{\bar\nu_e}$ for the different cases behave similarly to those in the energy luminosities shown in Fig.~\ref{fig:lumin} and discussed in the previous paragraphs. 
  
In order to further quantify the impact of weak magnetism and recoil contributions, we show in Fig.~\ref{fig:mfp_a} the density dependence of the inverse mean-free path (see Ref.~\cite{Fischer:2012a} for definition) for all neutrino flavors. The upper panel is for elastic neutral-current scattering on neutrons and protons, collectively denoted as $\nu N$, and the bottom panel is for inelastic processes, i.e. pair processes $\nu\bar\nu$ includes all of the reactions 8--10 in Table~\ref{tab:nu-reactions}, neutrino-electron/positron scattering $\nu e^\pm$, and charged-current $\nu_e n$-- and $\bar\nu_e p$--absorption. The conditions in Fig.~\ref{fig:mfp_a} correspond to the situation at about 5~s post bounce. The vertical lines in Fig.~\ref{fig:mfp_a} mark the locations of the neutrino spheres of last elastic (top panel) and inelastic (bottom panel) scattering. From Fig.~\ref{fig:mfp_a} it becomes clear that the elastic opacities for $\nu_e$ and $\nu_{\mu/\tau}$ are affected only marginally by the inclusion of weak magnetism and recoil; the opacity reduction due to weak magnetism and recoil is negligible for these neutrino flavors around their corresponding decoupling spheres. Note that without weak magnetism and recoil corrections to the $\bar\nu N$ scattering opacity, the location of the $\bar\nu_e$, $\nu_{\mu/\tau}$ and $\bar\nu_{\mu/\tau}$ spheres of last elastic scattering would coincide and the spectral differences between these flavors are only due to inelastic scattering on electrons/positrons, as well as the charged-current absorption for $\bar\nu_e$. Now, with weak magnetism and recoil corrections, the situation changes for antineutrinos. These corrections suppress the elastic $\bar\nu_e$ and $\bar\nu_{\mu/\tau}$ scattering opacity, which results in the shift of the neutrinospheres to higher densities and higher temperatures. Consequently, the average energies of the antineutrinos are enhanced, consistent with the discussion above and  Fig.~\ref{fig:lumin}. Low-density differences obtained when comparing the different treatments in Fig.~\ref{fig:mfp_a}, in particular for opacity channels involving $e^\pm$ and protons, are due to different $Y_e$, which will be further discussed in Sec.~\ref{sec:ejecta}. 

This is the first time that the enhancement of the average energies of $\bar\nu_e$ and $\bar\nu_{\mu/\tau}$ due to weak magnetism and recoil corrections is quantified. It was predicted in Ref.~\cite{Horowitz:2001xf} and observed in supernova simulations~\cite{MartinezPinedo:2012,Mirizzi:2016,Bollig:2017}. 

Let us now discuss the impact due to the inclusion of the inverse neutron decay as an additional opacity channel for $\bar\nu_e$ at low- and intermediate energies. While this new opacity channel has a negligible impact on the $\nu_e$ and $\nu_x$ ($\bar\nu_x$) fluxes and spectra during the early PNS deleptonization, the inclusion of the (inverse) neutron decay does result in a substantial reduction of the average energy for $\bar\nu_e$ (see Fig.~\ref{fig:lumin}). This result can be mainly traced to the low-energy enhancement of the charged-current $\bar\nu_e$--absorption opacity, as illustrated by the solid green lines for the {\em inelastic} $\bar\nu_e p$ process in the bottom panels of Figs.~\ref{fig:mfp_a} and \ref{fig:mfp_b}. The enhanced $\bar\nu_e p$--opacity shifts the neutrinosphere of last inelastic scattering to lower density, so $\bar\nu_e$ energetically decouple at lower temperatures with a lower average energy. However, the lower average $\bar\nu_e$ energy reduces the elastic opacity (top panels of Figs.~\ref{fig:mfp_a} and \ref{fig:mfp_b}), so paradoxically, the neutrinosphere of last elastic $\bar\nu_e$ scattering moves to higher density and temperature relative to cases without neutron decay. Nevertheless, the neutrinosphere of last elastic $\bar\nu_e$ scattering is always outside that of last inelastic $\bar\nu_e$ scattering. The above results also lead to slight reduction of the neutron abundance in the region of neutrino decoupling.

The enhanced $\bar\nu_e$--opacity at densities of $\simeq 10^{14}$~g~cm$^{-3}$ due to inverse neutron decay does not affect the PNS structure as long as charged-current absorption processes for both $\nu_e$ and $\bar\nu_e$ are treated with the full kinematics (see Fig.~\ref{fig:pns}). As mentioned above, the inclusion of the (inverse) neutron decay reduces the average $\bar\nu_e$ energy, and hence the {\em elastic} neutral-current neutrino-nucleon scattering opacity for $\bar\nu_e$ (see the upper panels of Figs.~\ref{fig:mfp_a} and \ref{fig:mfp_b}). On the other hand, the reduction of the neutrino-nucleon scattering opacity for $\nu_e$ in the region of decoupling is due to the higher $Y_e$ and hence lower neutron abundance. The same holds for the heavy-lepton flavors, but to a lesser extent due to their decoupling at slightly higher density. The reduced neutrino-neutron scattering inverse mean-free path results in the slight enhancement of the $\nu_e$ and $\bar\nu_e$ luminosities (see Fig.~\ref{fig:lumin}). The impact for the $\nu_x$ $(\bar\nu_x)$ is negligible because they decouple at somewhat higher density where the neutron abundance is not affected by the inclusion of the (inverse) neutron decay.  

\section{Impact on neutrino-driven wind nucleosynthesis}
\label{sec:ejecta}
The material at the PNS surface is subject to neutrino heating, mainly via reactions 1 and 2 in Table~\ref{tab:nu-reactions}, during the entire deleptonization phase. This leads to the subsequent ejection of the neutrino-driven wind, which is a site for the nucleosynthesis of heavy elements beyond the iron group. Note that the primary nucleosynthesis takes place during the early neutrino-driven wind phase when most mass is ejected.

The PNS evolution during the deleptonization, hence the nucleosynthesis of the neutrino-driven wind, is independent of the details of the explosion mechanism. Besides the properties of the remnant PNS, the nucleosynthesis conditions of the neutrino-driven wind are entirely determined by the neutrino fluxes and energies as well as their evolution during the PNS deleptonization~\cite{Qian:1996xt}. Larger (smaller) spectral differences between $\nu_e$ and $\bar\nu_e$ tend to result in more neutron-rich (proton-rich) conditions.

\begin{figure}[t!]
\includegraphics[width=\columnwidth]{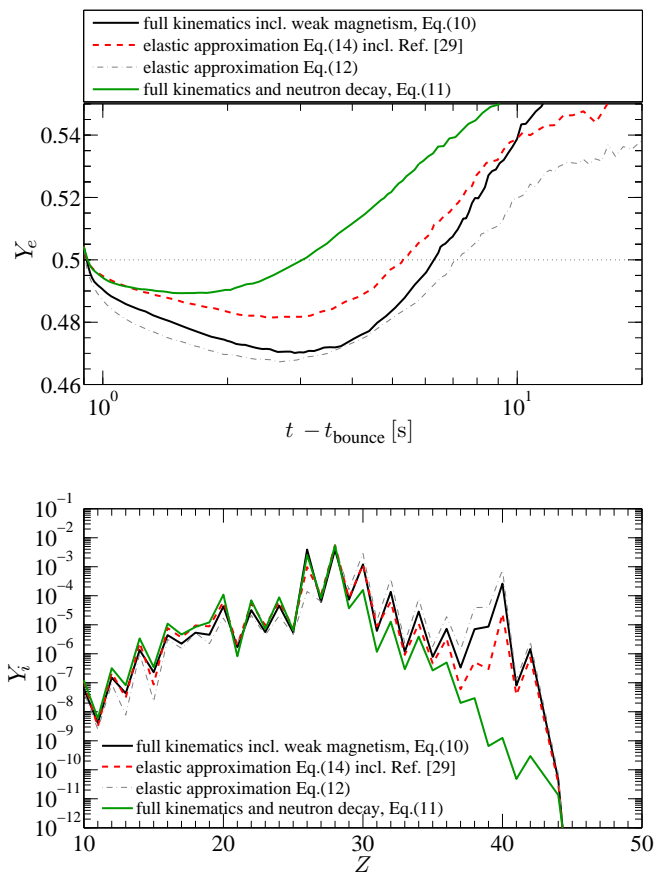}
\caption{(Color online) {\em Top panel}: 
Time evolution of $Y_e$ in the neutrino-driven wind (see text for detail), with four different treatments of the charged-current rates: full kinematics (black lines), elastic approximation including inelastic corrections and weak magnetism contributions (red lines)~\cite{Horowitz:2001xf}, elastic approximation without inelastic corrections and weak magnetism contributions (grey lines) and full kinematics including the neutron decay channel (green lines). {\em Bottom panel}: the corresponding integrated elemental nucleosynthesis results.}
\label{fig:ndw}
\end{figure}

The top panel of Fig.~\ref{fig:ndw} shows the time evolution of $Y_e$ of the neutrino-driven wind, obtained from detailed nucleosynthesis calculations~\cite{Wu:2015} which will be discussed further below. The $Y_e$ values correspond to the time when the temperature of the wind ejecta equals about $5\times 10^9$~K. Note that we do not include matter ejected by the supernova shock expansion. From Fig.~\ref{fig:ndw} it becomes clear that for the elastic charged-current rates indeed a slightly neutron-rich neutrino-driven wind is found during the early phase (grey dash-dotted line in Fig.~\ref{fig:ndw}). It corresponds to the phase with the largest spectral differences between $\nu_e$ and $\bar\nu_e$, which in turn are due to the medium effects via modified $Q$-values ($Q_0\pm\Delta U$) for the charged-current reactions 1 and 2 in Table~\ref{tab:nu-reactions}~(for details, c.f. Refs.~\cite{MartinezPinedo:2012,Roberts.Reddy.Shen:2012,GShen:2012}). They mainly suppress the $\bar\nu_e$-absorption opacity at low energies which shifts the average $\bar\nu_e$ energy to higher  values. The degree of neutron excess depends on the nuclear EOS. Towards later times, the spectral differences are reduced (see Fig.~\ref{fig:lumin}) and material turns proton rich. 

Now, the inclusion of weak magnetism corrections as well as nucleon recoil to the elastic rates, i.e. Eqs.~\eqref{eq:R} and \eqref{eq:Rbar}, as in Ref.~\cite{Horowitz:2001xf}, enhances the spectral differences between $\nu_e$ and $\bar\nu_e$ slightly (red lines in Fig.~\ref{fig:lumin}). Naively, one would expect the neutron excess to increase during the early neutrino-driven wind phase. However, this is not the case (red dashed line in Fig.~\ref{fig:ndw}). The neutron excess actually decreases because the modifications due to weak magnetism must also be applied to the reactions determining $Y_e$, which results in the reduction of the $\bar\nu_e$ cross section (see Fig.~\ref{fig1}). The reaction 2 in Table~\ref{tab:nu-reactions}, more precisely the inverse process, is consequently less efficient to maintain the moderately low $Y_e$ and material more readily turns proton rich (compare the grey dash-dotted and red dashed lines in Fig.~\ref{fig:ndw}).

With the full kinematics treatment of the charged-current processes for both $\bar\nu_e$ and $\nu_e$, the differences in their average energies and luminosities increase again (as discussed above, see the solid black lines in Fig.~\ref{fig:lumin}), which in turn increases the neutron excess back. However, with the inclusion of the inverse neutron decay as additional source of opacity for $\bar\nu_e$, the spectral difference between $\bar\nu_e$ and $\nu_e$ is reduced dramatically (see Fig.~\ref{fig:lumin}), turning material only slightly neutron rich with the lowest $Y_e\simeq 0.49$ during the early deleptonization between 1--2~s post bounce, while turning to the proton-rich side already at 3~s post bounce. This demonstrates how sensitive the nucleosynthesis conditions are to even small spectral changes from modifications of the neutrino opacity. 

We further calculate the corresponding nucleosynthesis yields in the neutrino-driven wind with these four different treatments of the charged-current weak processes using an established nuclear reaction network as in Refs.~\cite{Wu:2018mvg,Xiong2019}. Note that we have used the charged-current $\nu_e$ and $\bar\nu_e$ capture rates of Ref.~\cite{Horowitz:2001xf} in the reaction network to track the $Y_e$ evolution inside the wind, for the three cases other than the elastic approximation without any weak-magnetism and recoil corrections, as the three set of rates give consistent results in low density and temperature environment of the wind. The results are shown in the bottom panel of Fig.~\ref{fig:ndw}. Note again that these nucleosynthesis results contain the neutrino-driven wind only, i.e. explosive nucleosynthesis, which is associated with the shock expansion through the stellar envelope, is excluded.  With the inclusion of the inverse neutron decay and the lack of any significantly neutron-rich ejecta, the nucleosynthesis path terminates even before reaching the light neutron-capture elements associated with atomic numbers of $38<Z<42$, which had been reported previously based on the elastic treatment of weak rates due to the inclusion of medium modifications at the mean-field level \cite{MartinezPinedo:2012,MartinezPinedo:2014}.  However, material is also not proton-rich enough during the early evolution, when the neutrino fluxes are still substantial, such that we cannot obtain any $\nu p$ process \cite{Frohlich:2005ys}. This may change due to the convection inside the PNS (not included in this work) which tends to turn material to the proton-rich side at all times during the PNS deleptonization~\cite{Mirizzi:2016}. 

\section{Summary}
\label{sec:summary}
In this study we have explored a novel source of opacity in the charged-current absorption channel for $\bar\nu_e$ due to the inverse neutron decay. This process is enabled only due to medium modifications since the neutron and proton Fermi energies can be very different under supernova conditions. The medium modifications are treated at the mean-field level based on the nuclear RMF EOS, in a similar fashion for all charged-current absorption weak processes. The reaction rate for the (inverse) neutron decay is included in our spherically symmetric core collapse supernova model {\tt AGILE-BOLTZTRAN}. In contrast to previous studies~\cite{MartinezPinedo:2012,MartinezPinedo:2014}, here we develop the framework to account for inelastic contributions and weak magnetism corrections within the full kinematics framework for the electronic charged-current absorption reactions, while treating the neutral current neutrino nucleon scattering processes within the elastic approximation and accounting for these contributions following Ref.~\cite{Horowitz:2001xf}. 

 These corrections generally reduce the $\bar{\nu}_e$ cross sections which results in a slight increase of the corresponding luminosities and average energies~\cite{Liebendoerfer:2005a}. During the accretion phase, however, the inclusion of the (inverse) neutron decay leaves a negligible impact on the neutrino signal as well as on the dynamical evolution. This is associated with the physics of neutrino emission and heating/cooling, in other words neutrino decoupling, which is located in the thick layer of low-density material accumulated at the PNS surface during the accretion phase. There the medium modifications for the charged-current absorption processes are generally negligible, and hence the (inverse) neutron decay is suppressed.

The situation changes with the onset of the supernova explosion, when the stalled bounce shock is revived and moves to increasingly large radii. The subsequent evolution of the PNS is determined by the diffusion of neutrinos of all flavors and the consequent deleptonization of the PNS. During the first few 100~ms of the PNS deleptonization the thick layer of accumulated material at the PNS surface falls into the steep gravitational potential as mass accretion vanishes at the PNS surface. Consequently the neutrino fluxes are no longer determined by mass accretion but instead by diffusion. Neutrinos of all flavors decouple at generally higher density and temperature. The advantage of our spherically symmetric model lies in the inclusion of three-flavor Boltzmann neutrino transport. However, the details of the transition from accretion to diffusion dominated neutrino emission depend on the multidimensional nature of the proceeding supernova explosion. The associated delay may be enhanced when treated in multidimensional simulations~\cite{Bruenn:2014qea,Mueller:2012a,Mueller:2014,Suwa:2013}, despite difficulties reaching late-enough times beyond 1~s post bounce to capture the transition, and besides the approximate treatment of neutrino transport in such models. Moreover, it has been discussed that the supernova phenomena in general shows quantitatively different evolution in the full three-dimensional degrees of freedom~\cite{Lentz:2015,Melson:2015}.

During the PNS deleptonization phase it has been shown that $\bar\nu_e$ become increasingly similar to the heavy-lepton flavor neutrinos~\cite{Fischer:2012a}. It can be understood at the level of an analysis based on the neutrino opacity. The decoupling of $\bar\nu_e$ from matter is dominantly determined by neutral current elastic scattering on neutrons, while charged-current absorption on neutrons dominates the $\nu_e$-opacity. Here the (inverse) neutron decay is enabled in the $\bar\nu_e$-decoupling region due to the increased medium modifications at higher densities. The impact on the integrated neutrino observables is non-negligible, resulting in a substantial reduction of the average $\bar\nu_e$ energies, though obtained only with the fully inelastic treatment. The impact on the $\bar\nu_e$ spectra and fluxes is significantly smaller with the elastic approximation, which we confirmed by performing additional simulations employing the elastic rates for the (inverse) neutron decay only. In this paper we have demonstrated the importance of the correct treatment of inelastic contributions as well as weak magnetism with the full kinematics approach, in order to accurately capture the formation and evolution of the neutrino spectra.

This has important consequences for the nucleosynthesis conditions of the neutrino-driven wind ejected from the PNS surface during the deleptonization. Generally large differences are found from the inclusion of inelastic contributions and weak magnetism corrections within the full kinematics approach, in comparison to the elastic approximation including effectively inelastic contributions and weak magnetism corrections~\cite{Horowitz:2001xf}. Previously reported moderately large neutron excess, which included the medium modifications at the level of the elastic approximation and neglecting the inverse neutron decay~\cite{MartinezPinedo:2014}, cannot be confirmed. The full kinematics and inclusion of the inverse neutron decay reduce the neutron excess to a minimum of $Y_e\simeq 0.49$ during the early PNS deleptonization phase. Contributions from nuclear many-body correlations to both charged and neutral current processes~\cite{Horowitz.Caballero.ea:2017} may further alter spectral differences between $\bar\nu_e$ and $\nu_e$.

The present work demonstrates the importance of the consistent treatment of nuclear EOS and weak processes, opening channels that have not been considered before but can leave an impact on the neutrino fluxes and spectra. Moreover, this is essential for the prediction of reliable neutrino luminosities and spectra for potential observable signals from the next galactic supernova~\cite{Wu:2015}, as well as the nucleosynthesis conditions. Further progress will require the inclusion of nuclear many-body correlations as well as muonic weak processes~\cite{Bollig:2017}.

\begin{acknowledgments}
  The supernova simulations were performed at the Wroclaw Center for
  Scientific Computing and Networking (WCSS) in Wroclaw (Poland). TF
  acknowledges support from the Polish National Science Center (NCN)
  under grant number UMO-2016/23/B/ST2/00720.  GMP acknowledge the
  support of the Deutsche Forschungsgemeinschaft (DFG, German Research
  Foundation) - Projektnummer 279384907 - SFB 1245 ``Nuclei: From
  Fundamental Interactions to Structure and Stars''. MRW acknowledges
  support from the Ministry of Science and Technology, Taiwan under
  Grant No. 107-2119-M-001-038 and No. 108-2112-M-001-010. This work
  was supported by the COST Actions CA16117 ``ChETEC'' and CA16214
  ``PHAROS'', and by the US DOE grant DE-FG02-87ER40328.
\end{acknowledgments}

\begin{widetext}

\appendix
\section{expressions for $\Phi$}
\label{sec:X} 
In this section, we present the expression of $\Phi$ for $\nu_e$ ($\bar\nu_e$) capture. The expression without weak magnetism is obtained by setting $F_2 = 0$.

The matrix element in Eq.~(\ref{eq:m}) can be sorted into different four-momenta products as 
\begin{equation}
\label{eq:semi_amp_decomp}   
\begin{aligned}
 \left\langle |\mathcal{M}|^2\right\rangle
  = & ~A_{tot}M_A +B_{tot}M_B+ \cdots + K_{tot} M_K + L_{tot} M_L~ \\ 
 = & ~(4\,G)^2 \left[A_{tot}\left(p_1\cdot p_2^*\right)\left(p_3\cdot p_4^*\right)  \right. \\
 & +B_{tot}\left(p_1\cdot p_4^*\right)\left(p_3\cdot p_2^*\right) +C_{tot}\left(p_1\cdot p_2^*\right)^2\left(p_1\cdot p_3\right) \\
 & +D_{tot}\left(p_1\cdot p_2^*\right)\left(p_1\cdot p_3\right)^2 +E_{tot}\left(p_1\cdot p_2^*\right)^2 \\
 & +F_{tot}\left(p_1\cdot p_3\right)^2 +H_{tot}\left(p_1\cdot p_2^*\right)\left(p_1\cdot p_3\right) \\
 & +\left.J_{tot}\left(p_1\cdot p_2^*\right) +K_{tot}\left(p_1\cdot p_3\right) +L_{tot}\right]~.
\end{aligned}        
\end{equation}

Then the opacity can be expressed as 
\begin{equation}
\label{eq:opacity_red_wm2}
\chi(E_1) = \frac{G^2}{4\pi^3E_1^2}\int_{E_{3^-}}^{E_{3^+}}dE_3\left(1-f_3\right)\int_{E_{2^-}}^{E_{2^+}}dE_2 f_2\left(1-f_4\right) \Phi~,
\end{equation}
where
\begin{equation}
\begin{aligned}
\Phi &= \sum_X X_{tot} I_{X} \\
&= A_{tot}I_A+B_{tot}I_B+C_{tot}I_C+D_{tot}I_D+ E_{tot}I_E + F_{tot} I_F +H_{tot} I_H + J_{tot} I_J +K_{tot} I_K +L_{tot} I_L~, 
\label{eq:Xf1}
\end{aligned} 
\end{equation}
and
\begin{equation}
\begin{aligned}
\label{eq:I_x} 
 I_X &= \frac{\bar p_1 \bar p_2 \bar p_3 \bar p_4}{(2\pi)^2} \!\!\int\!\! d\Omega_2 d\Omega_3 d\Omega_4 dE_4 M_X  \delta^4\!\left(p_1+\!p_2-\!p_3-\!p_4\right) \\
 &=\frac{\bar p_1 \bar p_2 \bar p_3 \bar p_4}{(2\pi)^2} \!\!\int\!\! d\Omega_2 d\Omega_3 d\Omega_4 M_X \delta^3\!\left(\pmb{p}_1\!+\!\pmb{p}_2\!-\!\pmb{p}_3\!-\!\pmb{p}_4\right)~,
\end{aligned}  
\end{equation}
with $\bar p_i\equiv |{\bm p}_i|$. The coefficients $X_{tot}$ in Eq.~(\ref{eq:Xf1}) are 
\begin{equation}  
\begin{aligned} 
 & A_{tot}=\left(g_V\pm g_A\right)^2\pm 2g_AF_2\frac{m_2^*}{m_N}\Big(1-\frac{\Delta m^*}{2m_2^*}\Big), \\ & B_{tot}=\left(g_V\mp g_A\right)^2\mp 2g_AF_2\frac{m_2^*}{m_N}\Big(1-\frac{\Delta m^*}{2m_2^*}\Big)~, \\
 & C_{tot}=\frac{F_2^2}{m_N^2}, \\
 & D_{tot}=-\frac{F_2^2}{m_N^2}, \\ & E_{tot}=-\frac{F_2^2}{2m_N^2}[m_3^2-2\Delta U(E_3-E_1)+\Delta U^2]~, \\
 & F_{tot} = g_VF_2\frac{m_2^*}{m_N}\left(2-\frac{\Delta m_*}{m_2^*}\right) +\frac{F_2^2}{2m_N^2} \Big[ m_2^* m_4^* - Q + \frac{m_3^2}{4} - \Delta U(E_1 + E_2^*) - \frac{\Delta U^2}{4} \Big]~, \\
 & H_{tot} = \frac{F_2^2}{2m_N^2}\Big[ 2Q + m_3^2 + \Delta U(3E_1-E_3+2E_4^*) \Big], \\
 & J_{tot} = g_VF_2\frac{\Delta m^*}{2m_N} \left[m_3^2 - \Delta U(E_1+E_3)\right] + \frac{F_2^2}{2 m_N^{2}} J_{FF}~, \\      
 &  K_{tot} = \left(g_A^2-g_V^2\right)m_2^*m_4^* + g_V F_2 \frac{m_2^*}{2m_N} \Big\{ -3m_3^2 + 4\Delta U(E_3-E_1)-\Delta U^2 + \frac{\Delta m^*}{m_2^*} [2Q + m_3^2 + \Delta U(2E_1-E_3+E_4^*)] \Big\}  \\
 &\qquad+ \frac{F_2^2}{2m_N^2} K_{FF}~, \\   
 & L_{tot} =  g_VF_2\frac{m_2^*}{m_N} \Delta U E_1 \Big[ m_3^2 -\Delta U E_3 + \frac{\Delta m^*}{2m_2^*} \Big(-Q -\frac{m_3^2}{2}-\Delta U E_4^* +\frac{\Delta U^2}{2}\Big)\Big] + \frac{F_2^2}{2m_N^2}L_{FF}~, 
\end{aligned}
\end{equation} 
where the upper (lower) sign in `$\pm$' or `$\mp$' is for the neutrino (antineutrino) reaction, and 
\begin{equation}
\begin{aligned}
J_{FF} =& \Delta U \Big\{ -m_3^2\Big( E_1 + \frac{E_2^*+E_4^*}{2}\Big) + Q(E_3-3E_1) + \frac{\Delta U}{2} \big[ E_4^*(3E_3-5E_1) + E_2^*(E_3+E_1) + E_3^2-E_1^2 - 2Q\big] \\
& + \Delta U^2\Big( E_1 - E_3 - E_4^*\Big) + \frac{\Delta U^3}{2} \Big\}~, \\
K_{FF} =& -(m^*_2+3m^*_4)m^*_2 \frac{m_3^2}{4} + Q^2 + Q\frac{m_3^2}{4}-\frac{m_3^4}{8} \\ 
&+ \Delta U\Big[ \frac{Q}{2}(3E_1-E_2^*+E_3+3E_4^*)+\frac{m_3^2}{4}( 2E_2^* + E_3 + E_1) + m^*_2m^*_4(E_3-E_1) \Big]  \\
&+\Delta U^2 \Big[ \frac{1}{4}( m_2^{*2}-m^*_2m^*_4-3Q)  + \frac{E_4^*}{2} (E_3+2E_1-E_2^*+E_4^*) + E_2^*(\frac{1}{2}E_1-E_3) + \frac{E_1^2}{2}\Big] \\
& + \frac{\Delta U^3}{4} [ -E_1 + 2E_2^* - E_3 - 2E_4^*] + \frac{\Delta U^4}{8}~,  \\
L_{FF} = & \frac{\Delta U E_1}{4} \Big\{ m_3^2(m^*_2 + m^*_4)^2 - 4Q^2 + \Delta U[ -m_3^2( E_2^*+E_4^*+E_1) - 2E_3(m_2^{*2}+m^*_2m^*_4) + 2Q(E_2^*-3E_4^*-E_1) ]  \\
& + 2\Delta U^2[ E_2^*E_3 + E_4^*( E_2^*-E_4^*-E_1) + Q] + \Delta U^3 [  -E_2^* + E_4^* + E_1 ] \Big\}~,
\end{aligned}
\end{equation}
with $\Delta U=U_2-U_4$, $\Delta m^* = m_2^*-m_4^*$, and $Q = (m_2^{*2}-m_4^{*2})/2$. 

The integration over angles for $I_X$ in Eq.~(\ref{eq:I_x}) can be carried out analytically \cite{Lohs:2015}. The expressions for $I_{X=A,B,\cdots,L}$ are then given as follows,  
\begin{subequations}
\begin{equation}
\label{eq:IA}
 I_A=\frac{1}{60}  [3\left(p_{a+}^5-p_{a-}^5\right)-10\left(a+b\right)\left(p_{a+}^3-p_{a-}^3\right) + 60ab(p_{a+}-p_{a-})]~,
\end{equation}
\begin{equation}
\label{eq:IB}
 I_B=\frac{1}{60}  [3\left(p_{b+}^5-p_{b-}^5\right)-10\left(c+d\right)\left(p_{b+}^3-p_{b-}^3\right) + 60cd(p_{b+}-p_{b-})]~,
\end{equation}
\begin{equation}
\begin{aligned}
 I_C = & \Big[-\frac{\left(p_{a+}^7-p_{a-}^7\right)}{112}+\frac{a+\alpha_1}{20}\left(p_{a+}^5-p_{a-}^5\right) - \frac{a^2+4a\alpha_1-\alpha_0}{12}\left(p_{a+}^3-p_{a-}^3\right) \\
 + &\left(a^2\alpha_1-a\alpha_0\right)\left(p_{a+}-p_{a-}\right)-a^2\alpha_0\left(p_{a+}^{-1}-p_{a-}^{-1}\right)\Big]~, 
 \end{aligned}
\end{equation}
\begin{equation}
\begin{aligned}
\label{eq:SolID}
 I_D = & \Big[\frac{\left(p_{c+}^7-p_{c-}^7\right)}{112}+\frac{e+\epsilon_1}{20}\left(p_{c+}^5-p_{c-}^5\right) + \frac{e^2+4e\epsilon_1+\epsilon_0}{12}\left(p_{c+}^3-p_{c-}^3\right) \\
 + &\left(e^2\epsilon_1+e\epsilon_0\right)\left(p_{c+}-p_{c-}\right)-e^2\epsilon_0\left(p_{c+}^{-1}-p_{c-}^{-1}\right)\Big]~,
\end{aligned}
\end{equation}
\begin{equation}
\begin{aligned}
\label{eq:IE}
 I_E=\frac{1}{60} & \left[3\left(p_{a+}^5-p_{a-}^5\right)-20a\left(p_{a+}^3-p_{a-}^3\right) +60a^2\left(p_{a+}-p_{a-}\right)\right]~,
\end{aligned}
\end{equation}  
\begin{equation}
\label{eq:SolIF}
 I_F=\frac{1}{60} [3\left(p_{c+}^5-p_{c-}^5\right)+20e\left(p_{c+}^3-p_{c-}^3\right)+ 60e^2\left(p_{c+}-p_{c-}\right)], 
\end{equation}
\begin{equation}
I_H = [\frac{p_{c+}^5-p_{c-}^5}{40} +\frac{e+2\epsilon_1}{12}\left(p_{c+}^3-p_{c-}^3\right) + \frac{2e\epsilon_1+\epsilon_0}{2}\left(p_{c+}-p_{c-}\right)-e\epsilon_0\left(p_{c+}^{-1}-p_{c-}^{-1}\right)]~, 
\end{equation}
\begin{equation}
\label{eq:IJ}
I_J=\frac{1}{6} \left[-\left(p_{a+}^3-p_{a-}^3\right)+6a\left(p_{a+}-p_{a-}\right)\right]~ ,
\end{equation}
\begin{equation}
I_K = \frac{1}{6} \left[\left(p_{c+}^3-p_{c-}^3\right) +6e\left(p_{c+}-p_{c-}\right)\right]~,
\end{equation}
\begin{equation}
I_L = p_{a+}-p_{a-}~,
\end{equation}
\end{subequations} 
with 
\begin{equation}
\begin{aligned}
\label{eq:coeff_abcde}
& a =E_1E_2^* +\frac{\bar p_1^2+\bar p_2^2}{2}, \;\; b =E_3E_4^* +\frac{\bar p_3^2+\bar p_4^2}{2}~,  \\ 
& c = -E_1E_4^* +\frac{\bar p_1^2+\bar p_4^2}{2},\;\; d = -E_3E_2^* +\frac{\bar p_2^2+\bar p_3^2}{2}~,\;\; e =  E_1E_3-\frac{\bar p_1^2+\bar p_3^2}{2} , \\
& \alpha_0 =  \frac{1}{4}\left(\bar p_1^2-\bar p_2^2\right)\left(\bar p_4^2-\bar p_3^2\right)~,\;\; \alpha_1 =  E_1E_3-\frac{1}{4}\left(\bar p_1^2-\bar p_2^2+\bar p_3^2-\bar p_4^2\right), \\
& \epsilon_0 = \frac{1}{4}\left(\bar p_1^2-\bar p_3^2\right)\left(\bar p_2^2-\bar p_4^2\right)~,\;\; \epsilon_1 =  E_1E_2^*+\frac{1}{4}\left(\bar p_1^2+\bar p_2^2-\bar p_3^2-\bar p_4^2\right), 
\end{aligned}   
\end{equation}
and  
\begin{equation}
\begin{aligned}
p_{a-} = & \max\left\lbrace |\bar p_1-\bar p_2|,|\bar p_3-\bar p_4|\right\rbrace~, \;\; p_{a+} =  \min\left\lbrace \bar p_1+\bar p_2,\bar p_3+\bar p_4 \right\rbrace \\
p_{b-} = & \max\left\lbrace |\bar p_1-\bar p_4|,|\bar p_2-\bar p_3| \right\rbrace~, \;\; p_{b+} = \min\left\lbrace \bar p_1+\bar p_4,\bar p_2+\bar p_3 \right\rbrace, \\
p_{c-} = & \max\left\lbrace |\bar p_1-\bar p_3|,|\bar p_2-\bar p_4| \right\rbrace~, \;\;
p_{c+} =  \min\left\lbrace \bar p_1+\bar p_3,\bar p_2+\bar p_4 \right\rbrace~.
\end{aligned}
\end{equation}
In this study we determine the integration bounds of $E_{2,3}$ in Eq.~(\ref{eq:opacity_red_wm2}) numerically to ensure all regions are kinematically allowed. Another way is to set 
\begin{align}
& E_{2-}={\rm max}\{ m_2^*+U_2~, m_3+m_4^*+U_4-E_1\}~, \\
& E_{3-} = m_3, \;\; E_{3+}=E_1+E_2-m_4^*-U_4~,
\end{align}
from $E_{2,4}\ge m_{2,4}^*+U_{2,4}$ and $E_3\ge m_3$, and set the integrand to zero if $p_{a-} > p_{a+}$, $p_{b-} > p_{b+}$ or $p_{c-} > p_{c+}$.We notice that our treatment of weak magnetism corrections produces results that are numerically identical to those of Ref.~\cite{Roberts.Reddy:2017}.

\section{Relationship between the opacities of inverse neutron decay and neutrino absorption on neutrons}
\label{app:B}
Assuming non-relativistic nucleons, neglecting weak magnetism and the dependence of the matrix element on the angle between nucleons, the opacity for inverse neutron decay can be expressed as
\begin{equation}
\label{eq:invndecayopacity}
\chi(E_{\bar{\nu}_e})= \frac{G^2}{4\pi^2}
\left(g_V^2 + 3 g_A^2\right) \int_{m_e}^\infty dE_e
\frac{E_e}{E_{\bar{\nu}_e}} f(E_e) \int_{|E_{\bar{\nu}_e} -
  p_e|}^{E_{\bar{\nu}_e} + p_e} dq q \, S_{pn}(E_e+ E_{\bar{\nu}_e},q)~,
\end{equation}
where $S_{pn}(q_0,q)$ is the structure function that characterizes the isospin response of the nuclear medium to an energy transfer $q_0$ and a momentum transfer $q$~\cite{Reddy:1998}. It is interesting to compare this expression with the neutrino emissivity for the reaction $e^- + p \rightarrow n + \nu_e$~,
\begin{equation}
\label{eq:ecemis}
j(E_{\nu_e})= \frac{G^2}{4\pi^2}\left(g_V^2 + 3 g_A^2\right) \int_{m_e}^\infty dE_e \frac{E_e}{E_{\nu_e}} f(E_e) \int_{|E_{\nu_e} - p_e|}^{E_{\nu_e} + p_e} dq q \, S_{pn}(E_e-E_{\nu_e},q)~. 
\end{equation}
We expect very similar numerical values for both quantities due to the fact that both processes depend on the same response function and involve similar energy transfer, $q_0 \approx \Delta m_{np}^* + \Delta U_{np}$. Using the detailed balance relation
$\chi(E_{\nu_e}) = e^{\beta(E_{\nu_e} - \mu^{\text{eq}}_{\nu_e})}
j(E_{\nu_e})$, with
$\mu^{\text{eq}}_{\nu_e} = - \mu^{\text{eq}}_{\bar{\nu}_e} = \mu_e -
(\mu_n-\mu_p)$, we obtain for the opacity of
$\nu_e + n \rightarrow p + e^-$~,
\begin{equation}
\label{eq:nueopa}
\chi(E_{\nu_e})= \frac{G^2}{4\pi^2}\left(g_V^2 + 3 g_A^2\right) e^{\beta(E_{\nu_e}-\mu^{\text{eq}}_{\nu_e})}
\int_{m_e}^\infty dE_e \frac{E_e}{E_{\nu_e}}  f(E_e) \int_{|E_{\nu_e}
  - p_e|}^{E_{\nu_e} + p_e} dq q \, S_{pn}(E_e-E_{\nu_e},q).  
\end{equation}
For low (anti)neutrino energies the following approximate relation between the opacity for $\nu_e$ absorption and neutron decay is valid,
\begin{equation}
\label{eq:nuabsvsndec}
\chi_{\nu_e n \rightarrow p e^-}(E) \approx  e^{\beta(E-\mu^{\text{eq}}_{\nu_e})} \chi_{\bar\nu_e e^- p \rightarrow n} (E)~.
\end{equation}
\end{widetext}

%


\begin{thebibliography}{84}%
\makeatletter
\providecommand \@ifxundefined [1]{%
 \@ifx{#1\undefined}
}%
\providecommand \@ifnum [1]{%
 \ifnum #1\expandafter \@firstoftwo
 \else \expandafter \@secondoftwo
 \fi
}%
\providecommand \@ifx [1]{%
 \ifx #1\expandafter \@firstoftwo
 \else \expandafter \@secondoftwo
 \fi
}%
\providecommand \natexlab [1]{#1}%
\providecommand \enquote  [1]{``#1''}%
\providecommand \bibnamefont  [1]{#1}%
\providecommand \bibfnamefont [1]{#1}%
\providecommand \citenamefont [1]{#1}%
\providecommand \href@noop [0]{\@secondoftwo}%
\providecommand \href [0]{\begingroup \@sanitize@url \@href}%
\providecommand \@href[1]{\@@startlink{#1}\@@href}%
\providecommand \@@href[1]{\endgroup#1\@@endlink}%
\providecommand \@sanitize@url [0]{\catcode `\\12\catcode `\$12\catcode
  `\&12\catcode `\#12\catcode `\^12\catcode `\_12\catcode `\%12\relax}%
\providecommand \@@startlink[1]{}%
\providecommand \@@endlink[0]{}%
\providecommand \url  [0]{\begingroup\@sanitize@url \@url }%
\providecommand \@url [1]{\endgroup\@href {#1}{\urlprefix }}%
\providecommand \urlprefix  [0]{URL }%
\providecommand \Eprint [0]{\href }%
\providecommand \doibase [0]{http://dx.doi.org/}%
\providecommand \selectlanguage [0]{\@gobble}%
\providecommand \bibinfo  [0]{\@secondoftwo}%
\providecommand \bibfield  [0]{\@secondoftwo}%
\providecommand \translation [1]{[#1]}%
\providecommand \BibitemOpen [0]{}%
\providecommand \bibitemStop [0]{}%
\providecommand \bibitemNoStop [0]{.\EOS\space}%
\providecommand \EOS [0]{\spacefactor3000\relax}%
\providecommand \BibitemShut  [1]{\csname bibitem#1\endcsname}%
\let\auto@bib@innerbib\@empty
\bibitem [{\citenamefont {Bethe}\ and\ \citenamefont
  {Wilson}(1985)}]{Bethe:1985ux}%
  \BibitemOpen
  \bibfield  {author} {\bibinfo {author} {\bibfnamefont {H.~A.}\ \bibnamefont
  {Bethe}}\ and\ \bibinfo {author} {\bibfnamefont {R.}~\bibnamefont {Wilson},
  \bibfnamefont {James}},\ }\href {\doibase 10.1086/163343} {\bibfield
  {journal} {\bibinfo  {journal} {Astrophys.J.}\ }\textbf {\bibinfo {volume}
  {295}},\ \bibinfo {pages} {14} (\bibinfo {year} {1985})}\BibitemShut
  {NoStop}%
\bibitem [{\citenamefont {LeBlanc}\ and\ \citenamefont
  {Wilson}(1970)}]{LeBlanc:1970kg}%
  \BibitemOpen
  \bibfield  {author} {\bibinfo {author} {\bibfnamefont {J.~M.}\ \bibnamefont
  {LeBlanc}}\ and\ \bibinfo {author} {\bibfnamefont {J.~R.}\ \bibnamefont
  {Wilson}},\ }\href {\doibase 10.1086/150558} {\bibfield  {journal} {\bibinfo
  {journal} {Astrophys.J.}\ }\textbf {\bibinfo {volume} {161}},\ \bibinfo
  {pages} {541} (\bibinfo {year} {1970})}\BibitemShut {NoStop}%
\bibitem [{\citenamefont {Burrows}\ \emph {et~al.}(2006)\citenamefont
  {Burrows}, \citenamefont {Livne}, \citenamefont {Dessart}, \citenamefont
  {Ott},\ and\ \citenamefont {Murphy}}]{Burrows:2005dv}%
  \BibitemOpen
  \bibfield  {author} {\bibinfo {author} {\bibfnamefont {A.}~\bibnamefont
  {Burrows}}, \bibinfo {author} {\bibfnamefont {E.}~\bibnamefont {Livne}},
  \bibinfo {author} {\bibfnamefont {L.}~\bibnamefont {Dessart}}, \bibinfo
  {author} {\bibfnamefont {C.}~\bibnamefont {Ott}}, \ and\ \bibinfo {author}
  {\bibfnamefont {J.}~\bibnamefont {Murphy}},\ }\href {\doibase 10.1086/500174}
  {\bibfield  {journal} {\bibinfo  {journal} {Astrophys.J.}\ }\textbf {\bibinfo
  {volume} {640}},\ \bibinfo {pages} {878} (\bibinfo {year} {2006})}
  \BibitemShut {NoStop}%
\bibitem [{\citenamefont {Sagert}\ \emph {et~al.}(2009)\citenamefont {Sagert},
  \citenamefont {Fischer}, \citenamefont {Hempel}, \citenamefont {Pagliara},
  \citenamefont {Schaffner-Bielich} \emph {et~al.}}]{Sagert:2008ka}%
  \BibitemOpen
  \bibfield  {author} {\bibinfo {author} {\bibfnamefont {I.}~\bibnamefont
  {Sagert}}, \bibinfo {author} {\bibfnamefont {T.}~\bibnamefont {Fischer}},
  \bibinfo {author} {\bibfnamefont {M.}~\bibnamefont {Hempel}}, \bibinfo
  {author} {\bibfnamefont {G.}~\bibnamefont {Pagliara}}, \bibinfo {author}
  {\bibfnamefont {J.}~\bibnamefont {Schaffner-Bielich}},  \emph {et~al.},\
  }\href {\doibase 10.1103/PhysRevLett.102.081101} {\bibfield  {journal}
  {\bibinfo  {journal} {Phys.Rev.Lett.}\ }\textbf {\bibinfo {volume} {102}},\
  \bibinfo {pages} {081101} (\bibinfo {year} {2009})} \BibitemShut
  {NoStop}%
\bibitem [{\citenamefont {{Fischer}}\ \emph {et~al.}(2011)\citenamefont
  {{Fischer}}, \citenamefont {{Sagert}}, \citenamefont {{Pagliara}},
  \citenamefont {{Hempel}}, \citenamefont {{Schaffner-Bielich}}, \citenamefont
  {{Rauscher}}, \citenamefont {{Thielemann}}, \citenamefont {{K{\"a}ppeli}},
  \citenamefont {{Mart{\'{\i}}nez-Pinedo}},\ and\ \citenamefont
  {{Liebend{\"o}rfer}}}]{Fischer:2011}%
  \BibitemOpen
  \bibfield  {author} {\bibinfo {author} {\bibfnamefont {T.}~\bibnamefont
  {{Fischer}}}, \bibinfo {author} {\bibfnamefont {I.}~\bibnamefont {{Sagert}}},
  \bibinfo {author} {\bibfnamefont {G.}~\bibnamefont {{Pagliara}}}, \bibinfo
  {author} {\bibfnamefont {M.}~\bibnamefont {{Hempel}}}, \bibinfo {author}
  {\bibfnamefont {J.}~\bibnamefont {{Schaffner-Bielich}}}, \bibinfo {author}
  {\bibfnamefont {T.}~\bibnamefont {{Rauscher}}}, \bibinfo {author}
  {\bibfnamefont {F.-K.}\ \bibnamefont {{Thielemann}}}, \bibinfo {author}
  {\bibfnamefont {R.}~\bibnamefont {{K{\"a}ppeli}}}, \bibinfo {author}
  {\bibfnamefont {G.}~\bibnamefont {{Mart{\'{\i}}nez-Pinedo}}}, \ and\ \bibinfo
  {author} {\bibfnamefont {M.}~\bibnamefont {{Liebend{\"o}rfer}}},\ }\href
  {\doibase 10.1088/0067-0049/194/2/39} {\bibfield  {journal} {\bibinfo
  {journal} {\apjs}\ }\textbf {\bibinfo {volume} {194}},\ \bibinfo {eid} {39}
  (\bibinfo {year} {2011})} \BibitemShut {NoStop}%
\bibitem [{\citenamefont {{Fischer}}\ \emph {et~al.}(2018)\citenamefont
  {{Fischer}}, \citenamefont {{Bastian}}, \citenamefont {{Wu}}, \citenamefont
  {{Baklanov}}, \citenamefont {{Sorokina}}, \citenamefont {{Blinnikov}},
  \citenamefont {{Typel}}, \citenamefont {{Kl{\"a}hn}},\ and\ \citenamefont
  {{Blaschke}}}]{Fischer:2018}%
  \BibitemOpen
  \bibfield  {author} {\bibinfo {author} {\bibfnamefont {T.}~\bibnamefont
  {{Fischer}}}, \bibinfo {author} {\bibfnamefont {N.-U.~F.}\ \bibnamefont
  {{Bastian}}}, \bibinfo {author} {\bibfnamefont {M.-R.}\ \bibnamefont {{Wu}}},
  \bibinfo {author} {\bibfnamefont {P.}~\bibnamefont {{Baklanov}}}, \bibinfo
  {author} {\bibfnamefont {E.}~\bibnamefont {{Sorokina}}}, \bibinfo {author}
  {\bibfnamefont {S.}~\bibnamefont {{Blinnikov}}}, \bibinfo {author}
  {\bibfnamefont {S.}~\bibnamefont {{Typel}}}, \bibinfo {author} {\bibfnamefont
  {T.}~\bibnamefont {{Kl{\"a}hn}}}, \ and\ \bibinfo {author} {\bibfnamefont
  {D.~B.}\ \bibnamefont {{Blaschke}}},\ }\href {\doibase
  10.1038/s41550-018-0583-0} {\bibfield  {journal} {\bibinfo  {journal} {Nature
  Astronomy}\ }\textbf {\bibinfo {volume} {2}},\ \bibinfo {pages} {0583}
  (\bibinfo {year} {2018})} \BibitemShut {NoStop}%
\bibitem [{\citenamefont {{Janka}}\ \emph {et~al.}(2007)\citenamefont
  {{Janka}}, \citenamefont {{Langanke}}, \citenamefont {{Marek}}, \citenamefont
  {{Mart{\'{\i}}nez-Pinedo}},\ and\ \citenamefont {{M{\"u}ller}}}]{Janka:2007}%
  \BibitemOpen
  \bibfield  {author} {\bibinfo {author} {\bibfnamefont {H.-T.}\ \bibnamefont
  {{Janka}}}, \bibinfo {author} {\bibfnamefont {K.}~\bibnamefont {{Langanke}}},
  \bibinfo {author} {\bibfnamefont {A.}~\bibnamefont {{Marek}}}, \bibinfo
  {author} {\bibfnamefont {G.}~\bibnamefont {{Mart{\'{\i}}nez-Pinedo}}}, \ and\
  \bibinfo {author} {\bibfnamefont {B.}~\bibnamefont {{M{\"u}ller}}},\ }\href
  {\doibase 10.1016/j.physrep.2007.02.002} {\bibfield  {journal} {\bibinfo
  {journal} {\physrep}\ }\textbf {\bibinfo {volume} {442}},\ \bibinfo {pages}
  {38} (\bibinfo {year} {2007})}  \BibitemShut {NoStop}%
\bibitem [{\citenamefont {{Janka}}(2012)}]{Janka:2012}%
  \BibitemOpen
  \bibfield  {author} {\bibinfo {author} {\bibfnamefont {H.-T.}\ \bibnamefont
  {{Janka}}},\ }\href
  {\doibase 10.1146/annurev-nucl-102711-094901}
  {\bibfield {journal} {\bibinfo  {journal} {Ann. Rev. of Nucl. Part. Sci.}\ }
  \textbf {\bibinfo {volume} {62}},\ \bibinfo {pages} {407} (\bibinfo {year}{2012})}
  \BibitemShut {NoStop}%
\bibitem [{\citenamefont {{Janka}}\ \emph {et~al.}(2016)\citenamefont
  {{Janka}}, \citenamefont {{Melson}},\ and\ \citenamefont
  {{Summa}}}]{Janka.Melson.Summa:2016}%
  \BibitemOpen
  \bibfield  {author} {\bibinfo {author} {\bibfnamefont {H.-T.}\ \bibnamefont
  {{Janka}}}, \bibinfo {author} {\bibfnamefont {T.}~\bibnamefont {{Melson}}}, \
  and\ \bibinfo {author} {\bibfnamefont {A.}~\bibnamefont {{Summa}}},\ }\href
  {\doibase 10.1146/annurev-nucl-102115-044747}
  {\bibfield {journal} {\bibinfo  {journal} {Ann. Rev. of Nucl. Part. Sci.}\ }
  \textbf {\bibinfo {volume} {66}},\ \bibinfo {pages} {341} (\bibinfo {year} {2016})}
  \BibitemShut {NoStop}%
\bibitem [{\citenamefont {{Fischer}}\ \emph {et~al.}(2012)\citenamefont
  {{Fischer}}, \citenamefont {{Mart{\'{\i}}nez-Pinedo}}, \citenamefont
  {{Hempel}},\ and\ \citenamefont {{Liebend{\"o}rfer}}}]{Fischer:2012a}%
  \BibitemOpen
  \bibfield  {author} {\bibinfo {author} {\bibfnamefont {T.}~\bibnamefont
  {{Fischer}}}, \bibinfo {author} {\bibfnamefont {G.}~\bibnamefont
  {{Mart{\'{\i}}nez-Pinedo}}}, \bibinfo {author} {\bibfnamefont
  {M.}~\bibnamefont {{Hempel}}}, \ and\ \bibinfo {author} {\bibfnamefont
  {M.}~\bibnamefont {{Liebend{\"o}rfer}}},\ }\href {\doibase
  10.1103/PhysRevD.85.083003} {\bibfield  {journal} {\bibinfo  {journal}
  {\prd}\ }\textbf {\bibinfo {volume} {85}},\ \bibinfo {eid} {083003} (\bibinfo
  {year} {2012})}\BibitemShut {NoStop}%
\bibitem [{\citenamefont {{Wu}}\ \emph {et~al.}(2015)\citenamefont {{Wu}},
  \citenamefont {{Qian}}, \citenamefont {{Mart{\'{\i}}nez-Pinedo}},
  \citenamefont {{Fischer}},\ and\ \citenamefont {{Huther}}}]{Wu:2015}%
  \BibitemOpen
  \bibfield  {author} {\bibinfo {author} {\bibfnamefont {M.-R.}\ \bibnamefont
  {{Wu}}}, \bibinfo {author} {\bibfnamefont {Y.-Z.}\ \bibnamefont {{Qian}}},
  \bibinfo {author} {\bibfnamefont {G.}~\bibnamefont
  {{Mart{\'{\i}}nez-Pinedo}}}, \bibinfo {author} {\bibfnamefont
  {T.}~\bibnamefont {{Fischer}}}, \ and\ \bibinfo {author} {\bibfnamefont
  {L.}~\bibnamefont {{Huther}}},\ }\href {\doibase 10.1103/PhysRevD.91.065016}
  {\bibfield  {journal} {\bibinfo  {journal} {\prd}\ }\textbf {\bibinfo
  {volume} {91}},\ \bibinfo {eid} {065016} (\bibinfo {year} {2015})}\
   \BibitemShut{NoStop}%
\bibitem [{\citenamefont {{Fischer}}\ \emph {et~al.}(2016)\citenamefont
  {{Fischer}}, \citenamefont {{Chakraborty}}, \citenamefont {{Giannotti}},
  \citenamefont {{Mirizzi}}, \citenamefont {{Payez}},\ and\ \citenamefont
  {{Ringwald}}}]{Fischer:2016b}%
  \BibitemOpen
  \bibfield  {author} {\bibinfo {author} {\bibfnamefont {T.}~\bibnamefont
  {{Fischer}}}, \bibinfo {author} {\bibfnamefont {S.}~\bibnamefont
  {{Chakraborty}}}, \bibinfo {author} {\bibfnamefont {M.}~\bibnamefont
  {{Giannotti}}}, \bibinfo {author} {\bibfnamefont {A.}~\bibnamefont
  {{Mirizzi}}}, \bibinfo {author} {\bibfnamefont {A.}~\bibnamefont {{Payez}}},
  \ and\ \bibinfo {author} {\bibfnamefont {A.}~\bibnamefont {{Ringwald}}},\
  }\href {\doibase 10.1103/PhysRevD.94.085012} {\bibfield  {journal} {\bibinfo
  {journal} {\prd}\ }\textbf {\bibinfo {volume} {94}},\ \bibinfo {eid} {085012}
  (\bibinfo {year} {2016})}\
  \BibitemShut {NoStop}%
\bibitem [{\citenamefont {Pons}\ \emph {et~al.}(1999)\citenamefont {Pons},
  \citenamefont {Reddy}, \citenamefont {Prakash}, \citenamefont {Lattimer},\
  and\ \citenamefont {Miralles}}]{Pons:1998mm}%
  \BibitemOpen
  \bibfield  {author} {\bibinfo {author} {\bibfnamefont {J.}~\bibnamefont
  {Pons}}, \bibinfo {author} {\bibfnamefont {S.}~\bibnamefont {Reddy}},
  \bibinfo {author} {\bibfnamefont {M.}~\bibnamefont {Prakash}}, \bibinfo
  {author} {\bibfnamefont {J.}~\bibnamefont {Lattimer}}, \ and\ \bibinfo
  {author} {\bibfnamefont {J.}~\bibnamefont {Miralles}},\ }\href {\doibase
  10.1086/306889} {\bibfield  {journal} {\bibinfo  {journal} {Astrophys.J.}\
  }\textbf {\bibinfo {volume} {513}},\ \bibinfo {pages} {780} (\bibinfo {year}
  {1999})}\
  \BibitemShut {NoStop}%
\bibitem [{\citenamefont {{Page}}\ \emph {et~al.}(2000)\citenamefont {{Page}},
  \citenamefont {{Prakash}}, \citenamefont {{Lattimer}},\ and\ \citenamefont
  {{Steiner}}}]{Page:2000}%
  \BibitemOpen
  \bibfield  {author} {\bibinfo {author} {\bibfnamefont {D.}~\bibnamefont
  {{Page}}}, \bibinfo {author} {\bibfnamefont {M.}~\bibnamefont {{Prakash}}},
  \bibinfo {author} {\bibfnamefont {J.~M.}\ \bibnamefont {{Lattimer}}}, \ and\
  \bibinfo {author} {\bibfnamefont {A.~W.}\ \bibnamefont {{Steiner}}},\ }\href
  {\doibase 10.1103/PhysRevLett.85.2048} {\bibfield  {journal} {\bibinfo
  {journal} {Physical Review Letters}\ }\textbf {\bibinfo {volume} {85}},\
  \bibinfo {pages} {2048} (\bibinfo {year} {2000})},\
  \BibitemShut {NoStop}%
\bibitem [{\citenamefont {Fischer}\ \emph {et~al.}(2010)\citenamefont
  {Fischer}, \citenamefont {Whitehouse}, \citenamefont {Mezzacappa},
  \citenamefont {Thielemann},\ and\ \citenamefont
  {Liebend{\"o}rfer}}]{Fischer:2009af}%
  \BibitemOpen
  \bibfield  {author} {\bibinfo {author} {\bibfnamefont {T.}~\bibnamefont
  {Fischer}}, \bibinfo {author} {\bibfnamefont {S.}~\bibnamefont {Whitehouse}},
  \bibinfo {author} {\bibfnamefont {A.}~\bibnamefont {Mezzacappa}}, \bibinfo
  {author} {\bibfnamefont {F.-K.}\ \bibnamefont {Thielemann}}, \ and\ \bibinfo
  {author} {\bibfnamefont {M.}~\bibnamefont {Liebend{\"o}rfer}},\ }\href
  {\doibase 10.1051/0004-6361/200913106} {\bibfield  {journal} {\bibinfo
  {journal} {Astron.Astrophys.}\ }\textbf {\bibinfo {volume} {517}},\ \bibinfo
  {pages} {A80} (\bibinfo {year} {2010})}\
   \BibitemShut {NoStop}%
\bibitem [{\citenamefont {H{\"u}depohl}\ \emph {et~al.}(2010)\citenamefont
  {H{\"u}depohl}, \citenamefont {M{\"u}ller}, \citenamefont {Janka},
  \citenamefont {Marek},\ and\ \citenamefont {Raffelt}}]{Huedepohl:2010}%
  \BibitemOpen
  \bibfield  {author} {\bibinfo {author} {\bibfnamefont {L.}~\bibnamefont
  {H{\"u}depohl}}, \bibinfo {author} {\bibfnamefont {B.}~\bibnamefont
  {M{\"u}ller}}, \bibinfo {author} {\bibfnamefont {H.-T.}\ \bibnamefont
  {Janka}}, \bibinfo {author} {\bibfnamefont {A.}~\bibnamefont {Marek}}, \ and\
  \bibinfo {author} {\bibfnamefont {G.~G.}\ \bibnamefont {Raffelt}},\ }\href
  {\doibase 10.1103/PhysRevLett.104.251101} {\bibfield  {journal} {\bibinfo
  {journal} {Phys. Rev. Lett.}\ }\textbf {\bibinfo {volume} {104}},\ \bibinfo
  {pages} {251101} (\bibinfo {year} {2010})}\BibitemShut {NoStop}%
\bibitem [{\citenamefont {{Dasgupta}}\ \emph {et~al.}(2010)\citenamefont
  {{Dasgupta}}, \citenamefont {{Mirizzi}}, \citenamefont {{Tamborra}},\ and\
  \citenamefont {{Tom{\`a}s}}}]{Dasgupta:2010}%
  \BibitemOpen
  \bibfield  {author} {\bibinfo {author} {\bibfnamefont {B.}~\bibnamefont
  {{Dasgupta}}}, \bibinfo {author} {\bibfnamefont {A.}~\bibnamefont
  {{Mirizzi}}}, \bibinfo {author} {\bibfnamefont {I.}~\bibnamefont
  {{Tamborra}}}, \ and\ \bibinfo {author} {\bibfnamefont {R.}~\bibnamefont
  {{Tom{\`a}s}}},\ }\href {\doibase 10.1103/PhysRevD.81.093008} {\bibfield
  {journal} {\bibinfo  {journal} {\prd}\ }\textbf {\bibinfo {volume} {81}},\
  \bibinfo {eid} {093008} (\bibinfo {year} {2010})}\ 
  \BibitemShut {NoStop}%
\bibitem [{\citenamefont {{Serpico}}\ \emph {et~al.}(2012)\citenamefont
  {{Serpico}}, \citenamefont {{Chakraborty}}, \citenamefont {{Fischer}},
  \citenamefont {{H{\"u}depohl}}, \citenamefont {{Janka}},\ and\ \citenamefont
  {{Mirizzi}}}]{Serpico:2012}%
  \BibitemOpen
  \bibfield  {author} {\bibinfo {author} {\bibfnamefont {P.~D.}\ \bibnamefont
  {{Serpico}}}, \bibinfo {author} {\bibfnamefont {S.}~\bibnamefont
  {{Chakraborty}}}, \bibinfo {author} {\bibfnamefont {T.}~\bibnamefont
  {{Fischer}}}, \bibinfo {author} {\bibfnamefont {L.}~\bibnamefont
  {{H{\"u}depohl}}}, \bibinfo {author} {\bibfnamefont {H.-T.}\ \bibnamefont
  {{Janka}}}, \ and\ \bibinfo {author} {\bibfnamefont {A.}~\bibnamefont
  {{Mirizzi}}},\ }\href {\doibase 10.1103/PhysRevD.85.085031} {\bibfield
  {journal} {\bibinfo  {journal} {\prd}\ }\textbf {\bibinfo {volume} {85}},\
  \bibinfo {eid} {085031} (\bibinfo {year} {2012})}\ 
  \BibitemShut {NoStop}%
\bibitem [{\citenamefont {Reddy}\ \emph {et~al.}(1998)\citenamefont {Reddy},
  \citenamefont {Prakash},\ and\ \citenamefont {Lattimer}}]{Reddy:1998}%
  \BibitemOpen
  \bibfield  {author} {\bibinfo {author} {\bibfnamefont {S.}~\bibnamefont
  {Reddy}}, \bibinfo {author} {\bibfnamefont {M.}~\bibnamefont {Prakash}}, \
  and\ \bibinfo {author} {\bibfnamefont {J.~M.}\ \bibnamefont {Lattimer}},\
  }\href {\doibase 10.1103/PhysRevD.58.013009} {\bibfield  {journal} {\bibinfo
  {journal} {Phys. Rev. D}\ }\textbf {\bibinfo {volume} {58}},\ \bibinfo
  {pages} {013009} (\bibinfo {year} {1998})}\BibitemShut {NoStop}%
\bibitem [{\citenamefont {{Mart{\'{\i}}nez-Pinedo}}\ \emph
  {et~al.}(2012)\citenamefont {{Mart{\'{\i}}nez-Pinedo}}, \citenamefont
  {{Fischer}}, \citenamefont {{Lohs}},\ and\ \citenamefont
  {{Huther}}}]{MartinezPinedo:2012}%
  \BibitemOpen
  \bibfield  {author} {\bibinfo {author} {\bibfnamefont {G.}~\bibnamefont
  {{Mart{\'{\i}}nez-Pinedo}}}, \bibinfo {author} {\bibfnamefont
  {T.}~\bibnamefont {{Fischer}}}, \bibinfo {author} {\bibfnamefont
  {A.}~\bibnamefont {{Lohs}}}, \ and\ \bibinfo {author} {\bibfnamefont
  {L.}~\bibnamefont {{Huther}}},\ }\href {\doibase
  10.1103/PhysRevLett.109.251104} {\bibfield  {journal} {\bibinfo  {journal}
  {Phys. Rev. Lett.}\ }\textbf {\bibinfo {volume} {109}},\ \bibinfo {eid}
  {251104} (\bibinfo {year} {2012})}\BibitemShut {NoStop}%
\bibitem [{\citenamefont {{Roberts}}\ \emph {et~al.}(2012)\citenamefont
  {{Roberts}}, \citenamefont {{Reddy}},\ and\ \citenamefont
  {{Shen}}}]{Roberts.Reddy.Shen:2012}%
  \BibitemOpen
  \bibfield  {author} {\bibinfo {author} {\bibfnamefont {L.~F.}\ \bibnamefont
  {{Roberts}}}, \bibinfo {author} {\bibfnamefont {S.}~\bibnamefont {{Reddy}}},
  \ and\ \bibinfo {author} {\bibfnamefont {G.}~\bibnamefont {{Shen}}},\ }\href
  {\doibase 10.1103/PhysRevC.86.065803} {\bibfield  {journal} {\bibinfo
  {journal} {Phys. Rev. C}\ }\textbf {\bibinfo {volume} {86}},\ \bibinfo {eid}
  {065803} (\bibinfo {year} {2012})}\BibitemShut {NoStop}%
\bibitem [{\citenamefont {{Typel}}(2014)}]{Typel:2014}%
  \BibitemOpen
  \bibfield  {author} {\bibinfo {author} {\bibfnamefont {S.}~\bibnamefont
  {{Typel}}},\ }\href {\doibase 10.1103/PhysRevC.89.064321} {\bibfield
  {journal} {\bibinfo  {journal} {\prc}\ }\textbf {\bibinfo {volume} {89}},\
  \bibinfo {eid} {064321} (\bibinfo {year} {2014})},\ 
  \BibitemShut {NoStop}%
\bibitem [{\citenamefont {Roberts}\ \emph {et~al.}(2012)\citenamefont
  {Roberts}, \citenamefont {Shen}, \citenamefont {Cirigliano}, \citenamefont
  {Pons}, \citenamefont {Reddy},\ and\ \citenamefont
  {Woosley}}]{Roberts.Shen.ea:2012}%
  \BibitemOpen
  \bibfield  {author} {\bibinfo {author} {\bibfnamefont {L.~F.}\ \bibnamefont
  {Roberts}}, \bibinfo {author} {\bibfnamefont {G.}~\bibnamefont {Shen}},
  \bibinfo {author} {\bibfnamefont {V.}~\bibnamefont {Cirigliano}}, \bibinfo
  {author} {\bibfnamefont {J.~A.}\ \bibnamefont {Pons}}, \bibinfo {author}
  {\bibfnamefont {S.}~\bibnamefont {Reddy}}, \ and\ \bibinfo {author}
  {\bibfnamefont {S.~E.}\ \bibnamefont {Woosley}},\ }\href {\doibase
  10.1103/PhysRevLett.108.061103} {\bibfield  {journal} {\bibinfo  {journal}
  {Phys. Rev. Lett.}\ }\textbf {\bibinfo {volume} {108}},\ \bibinfo {pages}
  {061103} (\bibinfo {year} {2012})}\BibitemShut {NoStop}%
\bibitem [{\citenamefont {Woosley}\ \emph {et~al.}(1994)\citenamefont
  {Woosley}, \citenamefont {Wilson}, \citenamefont {Mathews}, \citenamefont
  {Hoffman},\ and\ \citenamefont {Meyer}}]{Woosley:1994ux}%
  \BibitemOpen
  \bibfield  {author} {\bibinfo {author} {\bibfnamefont {S.}~\bibnamefont
  {Woosley}}, \bibinfo {author} {\bibfnamefont {J.}~\bibnamefont {Wilson}},
  \bibinfo {author} {\bibfnamefont {G.}~\bibnamefont {Mathews}}, \bibinfo
  {author} {\bibfnamefont {R.}~\bibnamefont {Hoffman}}, \ and\ \bibinfo
  {author} {\bibfnamefont {B.}~\bibnamefont {Meyer}},\ }\href {\doibase
  10.1086/174638} {\bibfield  {journal} {\bibinfo  {journal} {Astrophys.J.}\
  }\textbf {\bibinfo {volume} {433}},\ \bibinfo {pages} {229} (\bibinfo {year}
  {1994})}\BibitemShut {NoStop}%
\bibitem [{\citenamefont {Takahashi}\ \emph {et~al.}(1994)\citenamefont
  {Takahashi}, \citenamefont {Witti},\ and\ \citenamefont
  {Janka}}]{Takahashi:1994yz}%
  \BibitemOpen
  \bibfield  {author} {\bibinfo {author} {\bibfnamefont {K.}~\bibnamefont
  {Takahashi}}, \bibinfo {author} {\bibfnamefont {J.}~\bibnamefont {Witti}}, \
  and\ \bibinfo {author} {\bibfnamefont {H.-T.}\ \bibnamefont {Janka}},\
  }\href@noop {} {\bibfield  {journal} {\bibinfo  {journal}
  {Astron.Astrophys.}\ }\textbf {\bibinfo {volume} {286}},\ \bibinfo {pages}
  {857} (\bibinfo {year} {1994})}\BibitemShut {NoStop}%
\bibitem [{\citenamefont {Qian}\ and\ \citenamefont
  {Woosley}(1996)}]{Qian:1996xt}%
  \BibitemOpen
  \bibfield  {author} {\bibinfo {author} {\bibfnamefont {Y.}~\bibnamefont
  {Qian}}\ and\ \bibinfo {author} {\bibfnamefont {S.}~\bibnamefont {Woosley}},\
  }\href {\doibase 10.1086/177973} {\bibfield  {journal} {\bibinfo  {journal}
  {Astrophys.J.}\ }\textbf {\bibinfo {volume} {471}},\ \bibinfo {pages} {331}
  (\bibinfo {year} {1996})},\ 
  \BibitemShut {NoStop}%
\bibitem [{\citenamefont {{Guo}}\ \emph {et~al.}()\citenamefont {{Guo}},
  \citenamefont {{Mart{\'{\i}}nez-Pinedo}},\ and\ \citenamefont
  {{Lohs}}}]{MartinezPinedo:2019}%
  \BibitemOpen
  \bibfield  {author} {\bibinfo {author} {\bibfnamefont {G.}~\bibnamefont
  {{Guo}}}, \bibinfo {author} {\bibfnamefont {G.}~\bibnamefont
  {{Mart{\'{\i}}nez-Pinedo}}}, \ and\ \bibinfo {author} {\bibfnamefont
  {A.}~\bibnamefont {{Lohs}}},\ }\href@noop {} {\bibinfo  {journal} {in
  preparation}\ }\BibitemShut {NoStop}%
\bibitem [{\citenamefont {Roberts}\ and\ \citenamefont
  {Reddy}(2017)}]{Roberts.Reddy:2017}%
  \BibitemOpen
\bibfield  {journal} {  }\bibfield  {author} {\bibinfo {author} {\bibfnamefont
  {L.~F.}\ \bibnamefont {Roberts}}\ and\ \bibinfo {author} {\bibfnamefont
  {S.}~\bibnamefont {Reddy}},\ }\href {\doibase 10.1103/PhysRevC.95.045807}
  {\bibfield  {journal} {\bibinfo  {journal} {Phys. Rev. C}\ }\textbf {\bibinfo
  {volume} {95}},\ \bibinfo {pages} {045807} (\bibinfo {year}
  {2017})}\BibitemShut {NoStop}%
\bibitem [{\citenamefont {Horowitz}(2002)}]{Horowitz:2001xf}%
  \BibitemOpen
  \bibfield  {author} {\bibinfo {author} {\bibfnamefont {C.}~\bibnamefont
  {Horowitz}},\ }\href {\doibase 10.1103/PhysRevD.65.043001} {\bibfield
  {journal} {\bibinfo  {journal} {Phys. Rev. D}\ }\textbf {\bibinfo {volume}
  {65}},\ \bibinfo {pages} {043001} (\bibinfo {year} {2002})}\BibitemShut
  {NoStop}%
\bibitem [{\citenamefont {Mart{\'{i}}nez-Pinedo}\ \emph
  {et~al.}(2014)\citenamefont {Mart{\'{i}}nez-Pinedo}, \citenamefont
  {Fischer},\ and\ \citenamefont {Huther}}]{MartinezPinedo:2014}%
  \BibitemOpen
  \bibfield  {author} {\bibinfo {author} {\bibfnamefont {G.}~\bibnamefont
  {Mart{\'{i}}nez-Pinedo}}, \bibinfo {author} {\bibfnamefont {T.}~\bibnamefont
  {Fischer}}, \ and\ \bibinfo {author} {\bibfnamefont {L.}~\bibnamefont
  {Huther}},\ }\href {\doibase 10.1088/0954-3899/41/4/044008} {\bibfield
  {journal} {\bibinfo  {journal} {J. Phys. G Nucl. Part. Phys.}\ }\textbf
  {\bibinfo {volume} {41}},\ \bibinfo {pages} {044008} (\bibinfo {year}
  {2014})}\BibitemShut {NoStop}%
\bibitem [{\citenamefont {Glashow}(1961)}]{Glashow:1961}%
  \BibitemOpen
  \bibfield  {author} {\bibinfo {author} {\bibfnamefont {S.~L.}\ \bibnamefont
  {Glashow}},\ }\href {\doibase 10.1016/0029-5582(61)90469-2} {\bibfield
  {journal} {\bibinfo  {journal} {Nucl. Phys.}\ }\textbf {\bibinfo {volume}
  {22}},\ \bibinfo {pages} {579 } (\bibinfo {year} {1961})}\BibitemShut
  {NoStop}%
\bibitem [{\citenamefont {Weinberg}(1967)}]{Weinberg:1967}%
  \BibitemOpen
  \bibfield  {author} {\bibinfo {author} {\bibfnamefont {S.}~\bibnamefont
  {Weinberg}},\ }\href {\doibase 10.1103/PhysRevLett.19.1264} {\bibfield
  {journal} {\bibinfo  {journal} {Phys. Rev. Lett.}\ }\textbf {\bibinfo
  {volume} {19}},\ \bibinfo {pages} {1264} (\bibinfo {year}
  {1967})}\BibitemShut {NoStop}%
\bibitem [{\citenamefont {Salam}(1968)}]{Salam:1968}%
  \BibitemOpen
  \bibfield  {author} {\bibinfo {author} {\bibfnamefont {A.}~\bibnamefont
  {Salam}},\ }in\ \href@noop {} {\emph {\bibinfo {booktitle} {Elementary
  Particle Physics: Relativistic Groups and Analyticity}}},\ \bibinfo {series
  and number} {\bibinfo {series} {Nobel Symposium}\ No.~\bibinfo {number}
  {8}},\ \bibinfo {editor} {edited by\ \bibinfo {editor} {\bibfnamefont
  {N.}~\bibnamefont {Svartholm}}}\ (\bibinfo  {publisher} {Almqvist and
  Wiksell, Stockholm},\ \bibinfo {year} {1968})\ p.\ \bibinfo {pages}
  {367}\BibitemShut {NoStop}%
\bibitem [{\citenamefont {Donnelly}\ and\ \citenamefont
  {Peccei}(1979)}]{Donnelly.Peccei:1979}%
  \BibitemOpen
  \bibfield  {author} {\bibinfo {author} {\bibfnamefont {T.~W.}\ \bibnamefont
  {Donnelly}}\ and\ \bibinfo {author} {\bibfnamefont {R.~P.}\ \bibnamefont
  {Peccei}},\ }\href {\doibase 10.1016/0370-1573(79)90010-3} {\bibfield
  {journal} {\bibinfo  {journal} {Phys. Repts.}\ }\textbf {\bibinfo {volume}
  {50}},\ \bibinfo {pages} {1} (\bibinfo {year} {1979})}\BibitemShut {NoStop}%
\bibitem [{\citenamefont {Walecka}(1975)}]{Walecka:1975}%
  \BibitemOpen
  \bibfield  {author} {\bibinfo {author} {\bibfnamefont {J.~D.}\ \bibnamefont
  {Walecka}},\ }in\ \href@noop {} {\emph {\bibinfo {booktitle} {Muon
  Physics}}},\ Vol.~\bibinfo {volume} {II},\ \bibinfo {editor} {edited by\
  \bibinfo {editor} {\bibfnamefont {V.~W.}\ \bibnamefont {Hughes}}\ and\
  \bibinfo {editor} {\bibfnamefont {C.~S.}\ \bibnamefont {Wu}}}\ (\bibinfo
  {publisher} {Academic Press},\ \bibinfo {address} {New York},\ \bibinfo
  {year} {1975})\ \bibinfo {type} {Section}\ \bibinfo {chapter} {V.4}, pp.\
  \bibinfo {pages} {113--218}\BibitemShut {NoStop}%
\bibitem [{\citenamefont {Tanabashi}\ \emph {et~al.}(2018)\citenamefont
  {Tanabashi} \emph {et~al.}}]{Tanabashi.Others:2018}%
  \BibitemOpen
  \bibfield  {author} {\bibinfo {author} {\bibfnamefont {M.}~\bibnamefont
  {Tanabashi}} \emph {et~al.} (\bibinfo {collaboration} {Particle Data
  Group}),\ }\href {\doibase 10.1103/PhysRevD.98.030001} {\bibfield  {journal}
  {\bibinfo  {journal} {Phys. Rev. D}\ }\textbf {\bibinfo {volume} {98}},\
  \bibinfo {pages} {030001} (\bibinfo {year} {2018})}\BibitemShut {NoStop}%
\bibitem [{\citenamefont {{Leinson}}\ and\ \citenamefont
  {{P{\'e}rez}}(2001)}]{Leinson.Perez:2001}%
  \BibitemOpen
  \bibfield  {author} {\bibinfo {author} {\bibfnamefont {L.~B.}\ \bibnamefont
  {{Leinson}}}\ and\ \bibinfo {author} {\bibfnamefont {A.}~\bibnamefont
  {{P{\'e}rez}}},\ }\href {\doibase 10.1016/S0370-2693(01)01042-5} {\bibfield
  {journal} {\bibinfo  {journal} {Phys. Lett. B}\ }\textbf {\bibinfo {volume}
  {518}},\ \bibinfo {pages} {15} (\bibinfo {year} {2001})}\BibitemShut
  {NoStop}%
\bibitem [{\citenamefont {{Leinson}}(2002)}]{Leinson:2002}%
  \BibitemOpen
  \bibfield  {author} {\bibinfo {author} {\bibfnamefont {L.~B.}\ \bibnamefont
  {{Leinson}}},\ }\href {\doibase 10.1016/S0375-9474(02)00991-0} {\bibfield
  {journal} {\bibinfo  {journal} {Nucl. Phys. A}\ }\textbf {\bibinfo {volume}
  {707}},\ \bibinfo {pages} {543} (\bibinfo {year} {2002})}\BibitemShut
  {NoStop}%
\bibitem [{\citenamefont {Lattimer}\ \emph {et~al.}(1991)\citenamefont
  {Lattimer}, \citenamefont {Pethick}, \citenamefont {Prakash},\ and\
  \citenamefont {Haensel}}]{Lattimer.Pethick.ea:1991}%
  \BibitemOpen
  \bibfield  {author} {\bibinfo {author} {\bibfnamefont {J.~M.}\ \bibnamefont
  {Lattimer}}, \bibinfo {author} {\bibfnamefont {C.~J.}\ \bibnamefont
  {Pethick}}, \bibinfo {author} {\bibfnamefont {M.}~\bibnamefont {Prakash}}, \
  and\ \bibinfo {author} {\bibfnamefont {P.}~\bibnamefont {Haensel}},\ }\href
  {\doibase 10.1103/PhysRevLett.66.2701} {\bibfield  {journal} {\bibinfo
  {journal} {Phys. Rev. Lett.}\ }\textbf {\bibinfo {volume} {66}},\ \bibinfo
  {pages} {2701} (\bibinfo {year} {1991})}\BibitemShut {NoStop}%
\bibitem [{\citenamefont {Yakovlev}\ and\ \citenamefont
  {Pethick}(2004)}]{Yakovlev.Pethick:2004}%
  \BibitemOpen
  \bibfield  {author} {\bibinfo {author} {\bibfnamefont {D.}~\bibnamefont
  {Yakovlev}}\ and\ \bibinfo {author} {\bibfnamefont {C.}~\bibnamefont
  {Pethick}},\ }\href {\doibase 10.1146/annurev.astro.42.053102.134013}
  {\bibfield  {journal} {\bibinfo  {journal} {Annu. Rev. Astron. Astrophys.}\
  }\textbf {\bibinfo {volume} {42}},\ \bibinfo {pages} {169} (\bibinfo {year}
  {2004})}\BibitemShut {NoStop}%
\bibitem [{\citenamefont {Bruenn}(1985)}]{Bruenn:1985en}%
  \BibitemOpen
  \bibfield  {author} {\bibinfo {author} {\bibfnamefont {S.~W.}\ \bibnamefont
  {Bruenn}},\ }\href {\doibase 10.1086/191056} {\bibfield  {journal} {\bibinfo
  {journal} {Astrophys.J.Suppl.}\ }\textbf {\bibinfo {volume} {58}},\ \bibinfo
  {pages} {771} (\bibinfo {year} {1985})}\BibitemShut {NoStop}%
\bibitem [{\citenamefont {Friman}\ and\ \citenamefont
  {Maxwell}(1979)}]{Friman.Maxwell:1979}%
  \BibitemOpen
  \bibfield  {author} {\bibinfo {author} {\bibfnamefont {B.~L.}\ \bibnamefont
  {Friman}}\ and\ \bibinfo {author} {\bibfnamefont {O.~V.}\ \bibnamefont
  {Maxwell}},\ }\href {\doibase 10.1086/157313} {\bibfield  {journal} {\bibinfo
   {journal} {Astrophys. J.}\ }\textbf {\bibinfo {volume} {232}},\ \bibinfo
  {pages} {541} (\bibinfo {year} {1979})}\BibitemShut {NoStop}%
\bibitem [{\citenamefont {Yakovlev}\ \emph {et~al.}(2001)\citenamefont
  {Yakovlev}, \citenamefont {Kaminker}, \citenamefont {Gnedin},\ and\
  \citenamefont {Haensel}}]{Yakovlev.Kaminker.ea:2001}%
  \BibitemOpen
  \bibfield  {author} {\bibinfo {author} {\bibfnamefont {D.}~\bibnamefont
  {Yakovlev}}, \bibinfo {author} {\bibfnamefont {A.~D.}\ \bibnamefont
  {Kaminker}}, \bibinfo {author} {\bibfnamefont {O.~Y.}\ \bibnamefont
  {Gnedin}}, \ and\ \bibinfo {author} {\bibfnamefont {P.}~\bibnamefont
  {Haensel}},\ }\href {\doibase 10.1016/S0370-1573(00)00131-9} {\bibfield
  {journal} {\bibinfo  {journal} {Phys. Rep.}\ }\textbf {\bibinfo {volume}
  {354}},\ \bibinfo {pages} {1} (\bibinfo {year} {2001})}\BibitemShut {NoStop}%
\bibitem [{\citenamefont {{Fischer}}\ \emph {et~al.}(2014)\citenamefont
  {{Fischer}}, \citenamefont {{Hempel}}, \citenamefont {{Sagert}},
  \citenamefont {{Suwa}},\ and\ \citenamefont
  {{Schaffner-Bielich}}}]{Fischer:2014}%
  \BibitemOpen
  \bibfield  {author} {\bibinfo {author} {\bibfnamefont {T.}~\bibnamefont
  {{Fischer}}}, \bibinfo {author} {\bibfnamefont {M.}~\bibnamefont {{Hempel}}},
  \bibinfo {author} {\bibfnamefont {I.}~\bibnamefont {{Sagert}}}, \bibinfo
  {author} {\bibfnamefont {Y.}~\bibnamefont {{Suwa}}}, \ and\ \bibinfo {author}
  {\bibfnamefont {J.}~\bibnamefont {{Schaffner-Bielich}}},\ }\href {\doibase
  10.1140/epja/i2014-14046-5} {\bibfield  {journal} {\bibinfo  {journal}
  {European Physical Journal A}\ }\textbf {\bibinfo {volume} {50}},\ \bibinfo
  {pages} {46} (\bibinfo {year} {2014})}\BibitemShut {NoStop}%
\bibitem [{\citenamefont {{Hempel}}(2015)}]{Hempel:2015a}%
  \BibitemOpen
  \bibfield  {author} {\bibinfo {author} {\bibfnamefont {M.}~\bibnamefont
  {{Hempel}}},\ }\href {\doibase 10.1103/PhysRevC.91.055807} {\bibfield
  {journal} {\bibinfo  {journal} {\prc}\ }\textbf {\bibinfo {volume} {91}},\
  \bibinfo {eid} {055807} (\bibinfo {year} {2015})},\
  \BibitemShut {NoStop}%
\bibitem [{\citenamefont {Liebend\"orfer}\ \emph {et~al.}(2001)\citenamefont
  {Liebend\"orfer}, \citenamefont {Mezzacappa},\ and\ \citenamefont
  {Thielemann}}]{Liebendoerfer:2001a}%
  \BibitemOpen
  \bibfield  {author} {\bibinfo {author} {\bibfnamefont {M.}~\bibnamefont
  {Liebend\"orfer}}, \bibinfo {author} {\bibfnamefont {A.}~\bibnamefont
  {Mezzacappa}}, \ and\ \bibinfo {author} {\bibfnamefont {F.-K.}\ \bibnamefont
  {Thielemann}},\ }\href {\doibase 10.1103/PhysRevD.63.104003} {\bibfield
  {journal} {\bibinfo  {journal} {Phys.Rev.}\ }\textbf {\bibinfo {volume}
  {D63}},\ \bibinfo {pages} {104003} (\bibinfo {year} {2001})}\BibitemShut
  {NoStop}%
\bibitem [{\citenamefont {Liebend{\"o}rfer}\ \emph {et~al.}(2001)\citenamefont
  {Liebend{\"o}rfer}, \citenamefont {Mezzacappa}, \citenamefont {Thielemann},
  \citenamefont {Messer}, \citenamefont {Hix} \emph{et~al.}}]{Liebendoerfer:2001b}%
  \BibitemOpen
  \bibfield  {author} {\bibinfo {author} {\bibfnamefont {M.}~\bibnamefont
  {Liebend{\"o}rfer}}, \bibinfo {author} {\bibfnamefont {A.}~\bibnamefont
  {Mezzacappa}}, \bibinfo {author} {\bibfnamefont {F.-K.}\ \bibnamefont
  {Thielemann}}, \bibinfo {author} {\bibfnamefont {O.~B.}\ \bibnamefont
  {Messer}}, \bibinfo {author} {\bibfnamefont {W.~R.}\ \bibnamefont {Hix}},
  \emph {et~al.},\ }\href {\doibase 10.1103/PhysRevD.63.103004} {\bibfield
  {journal} {\bibinfo  {journal} {Phys.Rev.}\ }\textbf {\bibinfo {volume}
  {D63}},\ \bibinfo {pages} {103004} (\bibinfo {year} {2001})}\
  \BibitemShut {NoStop}%
\bibitem [{\citenamefont {Liebend{\"o}rfer}\ \emph {et~al.}(2002)\citenamefont
  {Liebend{\"o}rfer}, \citenamefont {Rosswog},\ and\ \citenamefont
  {Thielemann}}]{Liebendoerfer:2002}%
  \BibitemOpen
  \bibfield  {author} {\bibinfo {author} {\bibfnamefont {M.}~\bibnamefont
  {Liebend{\"o}rfer}}, \bibinfo {author} {\bibfnamefont {S.}~\bibnamefont
  {Rosswog}}, \ and\ \bibinfo {author} {\bibfnamefont {F.-K.}\ \bibnamefont
  {Thielemann}},\ }\href {\doibase 10.1086/339872} {\bibfield  {journal}
  {\bibinfo  {journal} {Astrophys.J.Suppl.}\ }\textbf {\bibinfo {volume}
  {141}},\ \bibinfo {pages} {229} (\bibinfo {year} {2002})}\
  \BibitemShut {NoStop}%
\bibitem [{\citenamefont {Liebend\"orfer}\ \emph {et~al.}(2004)\citenamefont
  {Liebend\"orfer}, \citenamefont {Messer}, \citenamefont {Mezzacappa},
  \citenamefont {Bruenn}, \citenamefont {Cardall} \emph
  {et~al.}}]{Liebendoerfer:2004}%
  \BibitemOpen
  \bibfield  {author} {\bibinfo {author} {\bibfnamefont {M.}~\bibnamefont
  {Liebend\"orfer}}, \bibinfo {author} {\bibfnamefont {O.}~\bibnamefont
  {Messer}}, \bibinfo {author} {\bibfnamefont {A.}~\bibnamefont {Mezzacappa}},
  \bibinfo {author} {\bibfnamefont {S.}~\bibnamefont {Bruenn}}, \bibinfo
  {author} {\bibfnamefont {C.}~\bibnamefont {Cardall}},  \emph {et~al.},\
  }\href {\doibase 10.1086/380191} {\bibfield  {journal} {\bibinfo  {journal}
  {Astrophys.J.Suppl.}\ }\textbf {\bibinfo {volume} {150}},\ \bibinfo {pages}
  {263} (\bibinfo {year} {2004})}\ 
  \BibitemShut {NoStop}%
\bibitem [{\citenamefont {Liebend{\"o}rfer}\ \emph {et~al.}(2005)\citenamefont
  {Liebend{\"o}rfer}, \citenamefont {Rampp}, \citenamefont {Janka},\ and\
  \citenamefont {Mezzacappa}}]{Liebendoerfer:2005a}%
  \BibitemOpen
  \bibfield  {author} {\bibinfo {author} {\bibfnamefont {M.}~\bibnamefont
  {Liebend{\"o}rfer}}, \bibinfo {author} {\bibfnamefont {M.}~\bibnamefont
  {Rampp}}, \bibinfo {author} {\bibfnamefont {H.-T.}\ \bibnamefont {Janka}}, \
  and\ \bibinfo {author} {\bibfnamefont {A.}~\bibnamefont {Mezzacappa}},\
  }\href {\doibase 10.1086/427203} {\bibfield  {journal} {\bibinfo  {journal}
  {Astrophys.J.}\ }\textbf {\bibinfo {volume} {620}},\ \bibinfo {pages} {840}
  (\bibinfo {year} {2005})}\ 
  \BibitemShut {NoStop}%
\bibitem [{\citenamefont {Hempel}\ and\ \citenamefont
  {Schaffner-Bielich}(2010)}]{Hempel:2009mc}%
  \BibitemOpen
  \bibfield  {author} {\bibinfo {author} {\bibfnamefont {M.}~\bibnamefont
  {Hempel}}\ and\ \bibinfo {author} {\bibfnamefont {J.}~\bibnamefont
  {Schaffner-Bielich}},\ }\href {\doibase 10.1016/j.nuclphysa.2010.02.010}
  {\bibfield  {journal} {\bibinfo  {journal} {Nucl.Phys.}\ }\textbf {\bibinfo
  {volume} {A837}},\ \bibinfo {pages} {210} (\bibinfo {year} {2010})}\ 
  \BibitemShut {NoStop}%
\bibitem [{\citenamefont {{Typel}}(2005)}]{Typel:2005}%
  \BibitemOpen
  \bibfield  {author} {\bibinfo {author} {\bibfnamefont {S.}~\bibnamefont
  {{Typel}}},\ }\href {\doibase 10.1103/PhysRevC.71.064301} {\bibfield
  {journal} {\bibinfo  {journal} {\prc}\ }\textbf {\bibinfo {volume} {71}},\
  \bibinfo {eid} {064301} (\bibinfo {year} {2005})} \ 
  \BibitemShut {NoStop}%
\bibitem [{\citenamefont {Typel}\ \emph {et~al.}(2013)\citenamefont {Typel},
  \citenamefont {Oertel},\ and\ \citenamefont {Kl{\"a}hn}}]{Typel:2013rza}%
  \BibitemOpen
  \bibfield  {author} {\bibinfo {author} {\bibfnamefont {S.}~\bibnamefont
  {Typel}}, \bibinfo {author} {\bibfnamefont {M.}~\bibnamefont {Oertel}}, \
  and\ \bibinfo {author} {\bibfnamefont {T.}~\bibnamefont {Kl{\"a}hn}},\
  }\href@noop {} {\ \emph{eprint} arXiv astro-ph.SR/1307.5715 (\bibinfo {year} {2013})}\ 
  \BibitemShut {NoStop}%
\bibitem [{\citenamefont {Timmes}\ and\ \citenamefont
  {Arnett}(1999)}]{Timmes:1999}%
  \BibitemOpen
  \bibfield  {author} {\bibinfo {author} {\bibfnamefont {F.~X.}\ \bibnamefont
  {Timmes}}\ and\ \bibinfo {author} {\bibfnamefont {D.}~\bibnamefont
  {Arnett}},\ }\href
  {\doibase 10.1086/313271}
  {\bibfield  {journal} {\bibinfo  {journal}
  {\apjs.}\ }\textbf {\bibinfo {volume} {125}},\ \bibinfo {pages} {277} (\bibinfo {year} {1999})}
  \BibitemShut {NoStop}%
\bibitem [{\citenamefont {{Juodagalvis}}\ \emph {et~al.}(2010)\citenamefont
  {{Juodagalvis}}, \citenamefont {{Langanke}}, \citenamefont {{Hix}},
  \citenamefont {{Mart{\'{\i}}nez-Pinedo}},\ and\ \citenamefont
  {{Sampaio}}}]{Juodagalvis:2010}%
  \BibitemOpen
  \bibfield  {author} {\bibinfo {author} {\bibfnamefont {A.}~\bibnamefont
  {{Juodagalvis}}}, \bibinfo {author} {\bibfnamefont {K.}~\bibnamefont
  {{Langanke}}}, \bibinfo {author} {\bibfnamefont {W.~R.}\ \bibnamefont
  {{Hix}}}, \bibinfo {author} {\bibfnamefont {G.}~\bibnamefont
  {{Mart{\'{\i}}nez-Pinedo}}}, \ and\ \bibinfo {author} {\bibfnamefont {J.~M.}\
  \bibnamefont {{Sampaio}}},\ }\href 
  {\doibase 10.1016/j.nuclphysa.2010.09.012}
  {\bibfield  {journal} {\bibinfo  {journal} {Nuclear Physics A}\ }\textbf
  {\bibinfo {volume} {848}},\ \bibinfo {pages} {454} (\bibinfo {year}{2010})}\ 
  \BibitemShut {NoStop}%
\bibitem [{\citenamefont {Mezzacappa}\ and\ \citenamefont
  {Bruenn}(1993{\natexlab{a}})}]{Mezzacappa:1993gm}%
  \BibitemOpen
  \bibfield  {author} {\bibinfo {author} {\bibfnamefont {A.}~\bibnamefont
  {Mezzacappa}}\ and\ \bibinfo {author} {\bibfnamefont {S.}~\bibnamefont
  {Bruenn}},\ }\href {\doibase 10.1086/172394} {\bibfield  {journal} {\bibinfo
  {journal} {Astrophys.J.}\ }\textbf {\bibinfo {volume} {405}},\ \bibinfo
  {pages} {637} (\bibinfo {year} {1993}{\natexlab{a}})}\BibitemShut {NoStop}%
\bibitem [{\citenamefont {Mezzacappa}\ and\ \citenamefont
  {Bruenn}(1993{\natexlab{b}})}]{Mezzacappa:1993gx}%
  \BibitemOpen
  \bibfield  {author} {\bibinfo {author} {\bibfnamefont {A.}~\bibnamefont
  {Mezzacappa}}\ and\ \bibinfo {author} {\bibfnamefont {S.}~\bibnamefont
  {Bruenn}},\ }\href@noop {} {\bibfield  {journal} {\bibinfo  {journal}
  {Astrophys.J.}\ }\textbf {\bibinfo {volume} {410}},\ \bibinfo {pages} {740}
  (\bibinfo {year} {1993}{\natexlab{b}})}\BibitemShut {NoStop}%
\bibitem [{\citenamefont {Hannestad}\ and\ \citenamefont
  {Raffelt}(1998)}]{Hannestad:1997gc}%
  \BibitemOpen
  \bibfield  {author} {\bibinfo {author} {\bibfnamefont {S.}~\bibnamefont
  {Hannestad}}\ and\ \bibinfo {author} {\bibfnamefont {G.}~\bibnamefont
  {Raffelt}},\ }\href {\doibase 10.1086/306303} {\bibfield  {journal} {\bibinfo
   {journal} {Astrophys.J.}\ }\textbf {\bibinfo {volume} {507}},\ \bibinfo
  {pages} {339} (\bibinfo {year} {1998})}\ 
  \BibitemShut {NoStop}%
\bibitem [{\citenamefont {Buras}\ \emph {et~al.}(2003)\citenamefont {Buras},
  \citenamefont {Janka}, \citenamefont {Keil}, \citenamefont {Raffelt},\ and\
  \citenamefont {Rampp}}]{Buras:2002wt}%
  \BibitemOpen
  \bibfield  {author} {\bibinfo {author} {\bibfnamefont {R.}~\bibnamefont
  {Buras}}, \bibinfo {author} {\bibfnamefont {H.-T.}\ \bibnamefont {Janka}},
  \bibinfo {author} {\bibfnamefont {M.~T.}\ \bibnamefont {Keil}}, \bibinfo
  {author} {\bibfnamefont {G.~G.}\ \bibnamefont {Raffelt}}, \ and\ \bibinfo
  {author} {\bibfnamefont {M.}~\bibnamefont {Rampp}},\ }\href {\doibase
  10.1086/368015} {\bibfield  {journal} {\bibinfo  {journal} {Astrophys.J.}\
  }\textbf {\bibinfo {volume} {587}},\ \bibinfo {pages} {320} (\bibinfo {year} {2003})}\ 
  \BibitemShut {NoStop}%
\bibitem [{\citenamefont {{Fischer}}\ \emph {et~al.}(2009)\citenamefont
  {{Fischer}}, \citenamefont {{Whitehouse}}, \citenamefont {{Mezzacappa}},
  \citenamefont {{Thielemann}},\ and\ \citenamefont
  {{Liebend{\"o}rfer}}}]{Fischer:2009}%
  \BibitemOpen
  \bibfield  {author} {\bibinfo {author} {\bibfnamefont {T.}~\bibnamefont
  {{Fischer}}}, \bibinfo {author} {\bibfnamefont {S.~C.}\ \bibnamefont
  {{Whitehouse}}}, \bibinfo {author} {\bibfnamefont {A.}~\bibnamefont
  {{Mezzacappa}}}, \bibinfo {author} {\bibfnamefont {F.-K.}\ \bibnamefont
  {{Thielemann}}}, \ and\ \bibinfo {author} {\bibfnamefont {M.}~\bibnamefont
  {{Liebend{\"o}rfer}}},\ }\href {\doibase 10.1051/0004-6361/200811055}
  {\bibfield  {journal} {\bibinfo  {journal} {\aap}\ }\textbf {\bibinfo
  {volume} {499}},\ \bibinfo {pages} {1} (\bibinfo {year} {2009})}\ 
  \BibitemShut {NoStop}%
\bibitem [{\citenamefont {{Fuller}}\ and\ \citenamefont
  {{Meyer}}(1991)}]{Fuller:1991}%
  \BibitemOpen
  \bibfield  {author} {\bibinfo {author} {\bibfnamefont {G.~M.}\ \bibnamefont
  {{Fuller}}}\ and\ \bibinfo {author} {\bibfnamefont {B.~S.}\ \bibnamefont
  {{Meyer}}},\ }\href {\doibase 10.1086/170317} {\bibfield  {journal} {\bibinfo
   {journal} {\apj}\ }\textbf {\bibinfo {volume} {376}},\ \bibinfo {pages}
  {701} (\bibinfo {year} {1991})}\BibitemShut {NoStop}%
\bibitem [{\citenamefont {{Fischer}}\ \emph {et~al.}(2013)\citenamefont
  {{Fischer}}, \citenamefont {{Langanke}},\ and\ \citenamefont
  {{Mart{\'{\i}}nez-Pinedo}}}]{Fischer:2013}%
  \BibitemOpen
  \bibfield  {author} {\bibinfo {author} {\bibfnamefont {T.}~\bibnamefont
  {{Fischer}}}, \bibinfo {author} {\bibfnamefont {K.}~\bibnamefont
  {{Langanke}}}, \ and\ \bibinfo {author} {\bibfnamefont {G.}~\bibnamefont
  {{Mart{\'{\i}}nez-Pinedo}}},\ }\href {\doibase 10.1103/PhysRevC.88.065804}
  {\bibfield  {journal} {\bibinfo  {journal} {\prc}\ }\textbf {\bibinfo
  {volume} {88}},\ \bibinfo {eid} {065804} (\bibinfo {year}
  {2013})}\BibitemShut {NoStop}%
\bibitem [{\citenamefont {{Horowitz}}\ \emph
  {et~al.}(2012{\natexlab{a}})\citenamefont {{Horowitz}}, \citenamefont
  {{Shen}}, \citenamefont {{O'Connor}},\ and\ \citenamefont
  {{Ott}}}]{Horowitz:2012}%
  \BibitemOpen
  \bibfield  {author} {\bibinfo {author} {\bibfnamefont {C.~J.}\ \bibnamefont
  {{Horowitz}}}, \bibinfo {author} {\bibfnamefont {G.}~\bibnamefont {{Shen}}},
  \bibinfo {author} {\bibfnamefont {E.}~\bibnamefont {{O'Connor}}}, \ and\
  \bibinfo {author} {\bibfnamefont {C.~D.}\ \bibnamefont {{Ott}}},\ }\href
  {\doibase 10.1103/PhysRevC.86.065806} {\bibfield  {journal} {\bibinfo
  {journal} {Phys. Rev. C}\ }\textbf {\bibinfo {volume} {86}},\ \bibinfo {eid}
  {065806} (\bibinfo {year} {2012}{\natexlab{a}})}\BibitemShut {NoStop}%
\bibitem [{\citenamefont {Buras}\ \emph {et~al.}(2006)\citenamefont {Buras},
  \citenamefont {Rampp}, \citenamefont {Janka},\ and\ \citenamefont
  {Kifonidis}}]{Buras:2005rp}%
  \BibitemOpen
  \bibfield  {author} {\bibinfo {author} {\bibfnamefont {R.}~\bibnamefont
  {Buras}}, \bibinfo {author} {\bibfnamefont {M.}~\bibnamefont {Rampp}},
  \bibinfo {author} {\bibfnamefont {H.-T.}\ \bibnamefont {Janka}}, \ and\
  \bibinfo {author} {\bibfnamefont {K.}~\bibnamefont {Kifonidis}},\ }\href
  {\doibase 10.1051/0004-6361:20053783} {\bibfield  {journal} {\bibinfo
  {journal} {Astron.Astrophys.}\ }\textbf {\bibinfo {volume} {447}},\ \bibinfo
  {pages} {1049} (\bibinfo {year} {2006})},\ 
  \BibitemShut {NoStop}%
\bibitem [{\citenamefont {{Melson}}\ \emph {et~al.}(2015)\citenamefont
  {{Melson}}, \citenamefont {{Janka}},\ and\ \citenamefont
  {{Marek}}}]{Melson:2015}%
  \BibitemOpen
  \bibfield  {author} {\bibinfo {author} {\bibfnamefont {T.}~\bibnamefont
  {{Melson}}}, \bibinfo {author} {\bibfnamefont {H.-T.}\ \bibnamefont
  {{Janka}}}, \ and\ \bibinfo {author} {\bibfnamefont {A.}~\bibnamefont
  {{Marek}}},\ }\href {\doibase 10.1088/2041-8205/801/2/L24} {\bibfield
  {journal} {\bibinfo  {journal} {\apj}\ }\textbf {\bibinfo {volume} {801}},\
  \bibinfo {eid} {L24} (\bibinfo {year} {2015})}\
  \BibitemShut {NoStop}%
\bibitem [{\citenamefont {{Fischer}}\ \emph {et~al.}(2017)\citenamefont
  {{Fischer}}, \citenamefont {{Bastian}}, \citenamefont {{Blaschke}},
  \citenamefont {{Cierniak}}, \citenamefont {{Hempel}}, \citenamefont
  {{Kl{\"a}hn}}, \citenamefont {{Mart{\'{\i}}nez-Pinedo}}, \citenamefont
  {{Newton}}, \citenamefont {{R{\"o}pke}},\ and\ \citenamefont
  {{Typel}}}]{Fischer:2017}%
  \BibitemOpen
  \bibfield  {author} {\bibinfo {author} {\bibfnamefont {T.}~\bibnamefont
  {{Fischer}}}, \bibinfo {author} {\bibfnamefont {N.-U.}\ \bibnamefont
  {{Bastian}}}, \bibinfo {author} {\bibfnamefont {D.}~\bibnamefont
  {{Blaschke}}}, \bibinfo {author} {\bibfnamefont {M.}~\bibnamefont
  {{Cierniak}}}, \bibinfo {author} {\bibfnamefont {M.}~\bibnamefont
  {{Hempel}}}, \bibinfo {author} {\bibfnamefont {T.}~\bibnamefont
  {{Kl{\"a}hn}}}, \bibinfo {author} {\bibfnamefont {G.}~\bibnamefont
  {{Mart{\'{\i}}nez-Pinedo}}}, \bibinfo {author} {\bibfnamefont {W.~G.}\
  \bibnamefont {{Newton}}}, \bibinfo {author} {\bibfnamefont {G.}~\bibnamefont
  {{R{\"o}pke}}}, \ and\ \bibinfo {author} {\bibfnamefont {S.}~\bibnamefont
  {{Typel}}},\ }\href {\doibase 10.1017/pasa.2017.63} {\bibfield  {journal}
  {\bibinfo  {journal} {\pasa}\ }\textbf {\bibinfo {volume} {34}},\ \bibinfo
  {eid} {e067} (\bibinfo {year} {2017})}\ 
  \BibitemShut {NoStop}%
\bibitem [{\citenamefont {Woosley}\ \emph {et~al.}(2002)\citenamefont
  {Woosley}, \citenamefont {Heger},\ and\ \citenamefont
  {Weaver}}]{Woosley:2002zz}%
  \BibitemOpen
  \bibfield  {author} {\bibinfo {author} {\bibfnamefont {S.}~\bibnamefont
  {Woosley}}, \bibinfo {author} {\bibfnamefont {A.}~\bibnamefont {Heger}}, \
  and\ \bibinfo {author} {\bibfnamefont {T.}~\bibnamefont {Weaver}},\ }\href
  {\doibase 10.1103/RevModPhys.74.1015} {\bibfield  {journal} {\bibinfo
  {journal} {Rev.Mod.Phys.}\ }\textbf {\bibinfo {volume} {74}},\ \bibinfo
  {pages} {1015} (\bibinfo {year} {2002})}\BibitemShut {NoStop}%
\bibitem [{\citenamefont {{Fischer}}(2016{\natexlab{a}})}]{Fischer:2016c}%
  \BibitemOpen
  \bibfield  {author} {\bibinfo {author} {\bibfnamefont {T.}~\bibnamefont
  {{Fischer}}},\ }\href {\doibase 10.1140/epja/i2016-16054-9} {\bibfield
  {journal} {\bibinfo  {journal} {European Physical Journal A}\ }\textbf
  {\bibinfo {volume} {52}},\ \bibinfo {eid} {54} (\bibinfo {year}
  {2016}{\natexlab{a}})},\ 
  \BibitemShut {NoStop}%
\bibitem [{\citenamefont {{Fischer}}(2016{\natexlab{b}})}]{Fischer:2016a}%
  \BibitemOpen
  \bibfield  {author} {\bibinfo {author} {\bibfnamefont {T.}~\bibnamefont
  {{Fischer}}},\ }\href {\doibase 10.1051/0004-6361/201628991} {\bibfield
  {journal} {\bibinfo  {journal} {\aap}\ }\textbf {\bibinfo {volume} {593}},\
  \bibinfo {eid} {A103} (\bibinfo {year} {2016}{\natexlab{b}})},\ 
  \BibitemShut {NoStop}%
\bibitem [{\citenamefont {{Ugliano}}\ \emph {et~al.}(2012)\citenamefont
  {{Ugliano}}, \citenamefont {{Janka}}, \citenamefont {{Marek}},\ and\
  \citenamefont {{Arcones}}}]{Ugliano:2012}%
  \BibitemOpen
  \bibfield  {author} {\bibinfo {author} {\bibfnamefont {M.}~\bibnamefont
  {{Ugliano}}}, \bibinfo {author} {\bibfnamefont {H.-T.}\ \bibnamefont
  {{Janka}}}, \bibinfo {author} {\bibfnamefont {A.}~\bibnamefont {{Marek}}}, \
  and\ \bibinfo {author} {\bibfnamefont {A.}~\bibnamefont {{Arcones}}},\ }\href
  {\doibase 10.1088/0004-637X/757/1/69} {\bibfield  {journal} {\bibinfo
  {journal} {\apj}\ }\textbf {\bibinfo {volume} {757}},\ \bibinfo {eid} {69}
  (\bibinfo {year} {2012})}\ 
  \BibitemShut {NoStop}%
\bibitem [{\citenamefont {{Perego}}\ \emph {et~al.}(2015)\citenamefont
  {{Perego}}, \citenamefont {{Hempel}}, \citenamefont {{Fr{\"o}hlich}},
  \citenamefont {{Ebinger}}, \citenamefont {{Eichler}}, \citenamefont
  {{Casanova}}, \citenamefont {{Liebend{\"o}rfer}}\ and\ \citenamefont
  {{Thielemann}}}]{Perego:2015}%
  \BibitemOpen
  \bibfield  {author} {\bibinfo {author} {\bibfnamefont {A.}~\bibnamefont
  {{Perego}}}, \bibinfo {author} {\bibfnamefont {M.}~\bibnamefont {{Hempel}}},
  \bibinfo {author} {\bibfnamefont {C.}~\bibnamefont {{Fr{\"o}hlich}}},
  \bibinfo {author} {\bibfnamefont {K.}~\bibnamefont {{Ebinger}}}, \bibinfo
  {author} {\bibfnamefont {M.}~\bibnamefont {{Eichler}}}, \bibinfo {author}
  {\bibfnamefont {J.}~\bibnamefont {{Casanova}}}, \bibinfo {author}
  {\bibfnamefont {M.}~\bibnamefont {{Liebend{\"o}rfer}}}, \ and\ \bibinfo
  {author} {\bibfnamefont {F.-K.}\ \bibnamefont {{Thielemann}}},\ }\href
  {\doibase 10.1088/0004-637X/806/2/275} {\bibfield  {journal} {\bibinfo
  {journal} {\apj}\ }\textbf {\bibinfo {volume} {806}},\ \bibinfo {eid} {275}
  (\bibinfo {year} {2015})}\
  \BibitemShut {NoStop}%
\bibitem [{\citenamefont {{Mirizzi}}\ \emph {et~al.}(2016)\citenamefont
  {{Mirizzi}}, \citenamefont {{Tamborra}}, \citenamefont {{Janka}},
  \citenamefont {{Saviano}}, \citenamefont {{Scholberg}}, \citenamefont
  {{Bollig}}, \citenamefont {{H{\"u}depohl}},\ and\ \citenamefont
  {{Chakraborty}}}]{Mirizzi:2016}%
  \BibitemOpen
  \bibfield  {author} {\bibinfo {author} {\bibfnamefont {A.}~\bibnamefont
  {{Mirizzi}}}, \bibinfo {author} {\bibfnamefont {I.}~\bibnamefont
  {{Tamborra}}}, \bibinfo {author} {\bibfnamefont {H.-T.}\ \bibnamefont
  {{Janka}}}, \bibinfo {author} {\bibfnamefont {N.}~\bibnamefont {{Saviano}}},
  \bibinfo {author} {\bibfnamefont {K.}~\bibnamefont {{Scholberg}}}, \bibinfo
  {author} {\bibfnamefont {R.}~\bibnamefont {{Bollig}}}, \bibinfo {author}
  {\bibfnamefont {L.}~\bibnamefont {{H{\"u}depohl}}}, \ and\ \bibinfo {author}
  {\bibfnamefont {S.}~\bibnamefont {{Chakraborty}}},\ }\href {\doibase
  10.1393/ncr/i2016-10120-8} {\bibfield  {journal} {\bibinfo  {journal} {Nuovo
  Cimento Rivista Serie}\ }\textbf {\bibinfo {volume} {39}},\ \bibinfo {pages}
  {1} (\bibinfo {year} {2016})}\
  \BibitemShut {NoStop}%
\bibitem [{\citenamefont {{Bollig}}\ \emph {et~al.}(2017)\citenamefont
  {{Bollig}}, \citenamefont {{Janka}}, \citenamefont {{Lohs}}, \citenamefont
  {{Mart{\'{\i}}nez-Pinedo}}, \citenamefont {{Horowitz}},\ and\ \citenamefont
  {{Melson}}}]{Bollig:2017}%
  \BibitemOpen
  \bibfield  {author} {\bibinfo {author} {\bibfnamefont {R.}~\bibnamefont
  {{Bollig}}}, \bibinfo {author} {\bibfnamefont {H.-T.}\ \bibnamefont
  {{Janka}}}, \bibinfo {author} {\bibfnamefont {A.}~\bibnamefont {{Lohs}}},
  \bibinfo {author} {\bibfnamefont {G.}~\bibnamefont
  {{Mart{\'{\i}}nez-Pinedo}}}, \bibinfo {author} {\bibfnamefont {C.~J.}\
  \bibnamefont {{Horowitz}}}, \ and\ \bibinfo {author} {\bibfnamefont
  {T.}~\bibnamefont {{Melson}}},\ }\href {\doibase
  10.1103/PhysRevLett.119.242702} {\bibfield  {journal} {\bibinfo  {journal}
  {Physical Review Letters}\ }\textbf {\bibinfo {volume} {119}},\ \bibinfo
  {eid} {242702} (\bibinfo {year} {2017})}\
  \BibitemShut {NoStop}%
\bibitem [{\citenamefont {{Horowitz}}\ \emph
  {et~al.}(2012{\natexlab{b}})\citenamefont {{Horowitz}}, \citenamefont
  {{Shen}}, \citenamefont {{O'Connor}},\ and\ \citenamefont
  {{Ott}}}]{GShen:2012}%
  \BibitemOpen
  \bibfield  {author} {\bibinfo {author} {\bibfnamefont {C.~J.}\ \bibnamefont
  {{Horowitz}}}, \bibinfo {author} {\bibfnamefont {G.}~\bibnamefont {{Shen}}},
  \bibinfo {author} {\bibfnamefont {E.}~\bibnamefont {{O'Connor}}}, \ and\
  \bibinfo {author} {\bibfnamefont {C.~D.}\ \bibnamefont {{Ott}}},\ }\href
  {\doibase 10.1103/PhysRevC.86.065806} {\bibfield  {journal} {\bibinfo
  {journal} {\prc}\ }\textbf {\bibinfo {volume} {86}},\ \bibinfo {eid} {065806}
  (\bibinfo {year} {2012}{\natexlab{b}})}\ 
  \BibitemShut {NoStop}%
\bibitem [{\citenamefont {Wu}\ \emph {et~al.}(2019)\citenamefont {Wu},
  \citenamefont {Barnes}, \citenamefont {Martinez-Pinedo},\ and\ \citenamefont
  {Metzger}}]{Wu:2018mvg}%
  \BibitemOpen
  \bibfield  {author} {\bibinfo {author} {\bibfnamefont {M.-R.}\ \bibnamefont
  {Wu}}, \bibinfo {author} {\bibfnamefont {J.}~\bibnamefont {Barnes}}, \bibinfo
  {author} {\bibfnamefont {G.}~\bibnamefont {Martinez-Pinedo}}, \ and\ \bibinfo
  {author} {\bibfnamefont {B.~D.}\ \bibnamefont {Metzger}},\ }\href {\doibase
  10.1103/PhysRevLett.122.062701} {\bibfield  {journal} {\bibinfo  {journal}
  {Phys. Rev. Lett.}\ }\textbf {\bibinfo {volume} {122}},\ \bibinfo {pages}
  {062701} (\bibinfo {year} {2019})}\ 
  \BibitemShut {NoStop}%
\bibitem [{\citenamefont {{Xiong}}\ \emph {et~al.}(2019)\citenamefont
  {{Xiong}}, \citenamefont {{Wu}},\ and\ \citenamefont {{Qian}}}]{Xiong2019}%
  \BibitemOpen
  \bibfield  {author} {\bibinfo {author} {\bibfnamefont {Z.}~\bibnamefont
  {{Xiong}}}, \bibinfo {author} {\bibfnamefont {M.-R.}\ \bibnamefont {{Wu}}}, \
  and\ \bibinfo {author} {\bibfnamefont {Y.-Z.}\ \bibnamefont {{Qian}}},\
  }\href {\doibase 10.3847/1538-4357/ab2870} {\bibfield  {journal} {\bibinfo
  {journal} {\apj}\ }\textbf {\bibinfo {volume} {880}},\ \bibinfo {eid} {81}
  (\bibinfo {year} {2019})}\
  \BibitemShut {NoStop}%
\bibitem [{\citenamefont {Fr{\"o}hlich}\ \emph {et~al.}(2006)\citenamefont
  {Frohlich}, \citenamefont {Martinez-Pinedo}, \citenamefont {Liebend{\"o}rfer},
  \citenamefont {Thielemann}, \citenamefont {Bravo} \emph
  {et~al.}}]{Frohlich:2005ys}%
  \BibitemOpen
  \bibfield  {author} {\bibinfo {author} {\bibfnamefont {C.}~\bibnamefont
  {Frohlich}}, \bibinfo {author} {\bibfnamefont {G.}~\bibnamefont
  {Martinez-P{\'i}nedo}}, \bibinfo {author} {\bibfnamefont {M.}~\bibnamefont
  {Liebend{\"o}rfer}}, \bibinfo {author} {\bibfnamefont {F.-K.}\ \bibnamefont
  {Thielemann}}, \bibinfo {author} {\bibfnamefont {E.}~\bibnamefont {Bravo}},
  \emph {et~al.},\ }\href {\doibase 10.1103/PhysRevLett.96.142502} {\bibfield
  {journal} {\bibinfo  {journal} {Phys.Rev.Lett.}\ }\textbf {\bibinfo {volume}
  {96}},\ \bibinfo {pages} {142502} (\bibinfo {year} {2006})}\
  \BibitemShut {NoStop}%
\bibitem [{\citenamefont {Bruenn}\ \emph {et~al.}(2016)\citenamefont {Bruenn}
  \emph {et~al.}}]{Bruenn:2014qea}%
  \BibitemOpen
  \bibfield  {author} {\bibinfo {author} {\bibfnamefont {S.~W.}\ \bibnamefont
  {Bruenn}} \emph {et~al.},\ }\href {\doibase 10.3847/0004-637X/818/2/123}
  {\bibfield  {journal} {\bibinfo  {journal} {Astrophys. J.}\ }\textbf
  {\bibinfo {volume} {818}},\ \bibinfo {pages} {123} (\bibinfo {year}
  {2016})}\
  \BibitemShut {NoStop}%
\bibitem [{\citenamefont {{M{\"u}ller}}\ \emph {et~al.}(2012)\citenamefont
  {{M{\"u}ller}}, \citenamefont {{Janka}},\ and\ \citenamefont
  {{Marek}}}]{Mueller:2012a}%
  \BibitemOpen
  \bibfield  {author} {\bibinfo {author} {\bibfnamefont {B.}~\bibnamefont
  {{M{\"u}ller}}}, \bibinfo {author} {\bibfnamefont {H.-T.}\ \bibnamefont
  {{Janka}}}, \ and\ \bibinfo {author} {\bibfnamefont {A.}~\bibnamefont
  {{Marek}}},\ }\href
  {\doibase 10.1088/0004-637X/756/1/84}
  {\bibfield {journal} {\bibinfo  {journal} {\apj}\ }\textbf {\bibinfo {volume} {756}},\
  \bibinfo {eid} {84} (\bibinfo {year} {2012})}
  \BibitemShut {NoStop}%
\bibitem [{\citenamefont {{M{\"u}ller}}\ and\ \citenamefont
  {{Janka}}(2014)}]{Mueller:2014}%
  \BibitemOpen
  \bibfield  {author} {\bibinfo {author} {\bibfnamefont {B.}~\bibnamefont
  {{M{\"u}ller}}}\ and\ \bibinfo {author} {\bibfnamefont {H.-T.}\ \bibnamefont
  {{Janka}}},\ }\href
  {\doibase  10.1088/0004-637X/788/1/82}
  {\bibfield {journal} {\bibinfo  {journal} {\apj}\ }\textbf {\bibinfo {volume} {788}},\
  \bibinfo {eid} {82} (\bibinfo {year} {2014})}  
  \BibitemShut {NoStop}%
\bibitem [{\citenamefont {{Suwa}}\ \emph {et~al.}(2013)\citenamefont {{Suwa}},
  \citenamefont {{Takiwaki}}, \citenamefont {{Kotake}}, \citenamefont
  {{Fischer}}, \citenamefont {{Liebend{\"o}rfer}},\ and\ \citenamefont
  {{Sato}}}]{Suwa:2013}%
  \BibitemOpen
  \bibfield  {author} {\bibinfo {author} {\bibfnamefont {Y.}~\bibnamefont
  {{Suwa}}}, \bibinfo {author} {\bibfnamefont {T.}~\bibnamefont {{Takiwaki}}},
  \bibinfo {author} {\bibfnamefont {K.}~\bibnamefont {{Kotake}}}, \bibinfo
  {author} {\bibfnamefont {T.}~\bibnamefont {{Fischer}}}, \bibinfo {author}
  {\bibfnamefont {M.}~\bibnamefont {{Liebend{\"o}rfer}}}, \ and\ \bibinfo
  {author} {\bibfnamefont {K.}~\bibnamefont {{Sato}}},\ }\href
  {\doibase 10.1088/0004-637X/764/1/99} 
  {\bibfield  {journal} {\bibinfo  {journal} {\apj}\ }\textbf {\bibinfo {volume} {764}},\ \bibinfo {eid} {99} (\bibinfo {year} {2013})}\
  \BibitemShut {NoStop}%
\bibitem [{\citenamefont {{Lentz}}\ \emph {et~al.}(2015)\citenamefont
  {{Lentz}}, \citenamefont {{Bruenn}}, \citenamefont {{Hix}}, \citenamefont
  {{Mezzacappa}}, \citenamefont {{Messer}}, \citenamefont {{Endeve}},
  \citenamefont {{Blondin}}, \citenamefont {{Harris}}, \citenamefont
  {{Marronetti}},\ and\ \citenamefont {{Yakunin}}}]{Lentz:2015}%
  \BibitemOpen
  \bibfield  {author} {\bibinfo {author} {\bibfnamefont {E.~J.}\ \bibnamefont
  {{Lentz}}}, \bibinfo {author} {\bibfnamefont {S.~W.}\ \bibnamefont
  {{Bruenn}}}, \bibinfo {author} {\bibfnamefont {W.~R.}\ \bibnamefont {{Hix}}},
  \bibinfo {author} {\bibfnamefont {A.}~\bibnamefont {{Mezzacappa}}}, \bibinfo
  {author} {\bibfnamefont {O.~E.~B.}\ \bibnamefont {{Messer}}}, \bibinfo
  {author} {\bibfnamefont {E.}~\bibnamefont {{Endeve}}}, \bibinfo {author}
  {\bibfnamefont {J.~M.}\ \bibnamefont {{Blondin}}}, \bibinfo {author}
  {\bibfnamefont {J.~A.}\ \bibnamefont {{Harris}}}, \bibinfo {author}
  {\bibfnamefont {P.}~\bibnamefont {{Marronetti}}}, \ and\ \bibinfo {author}
  {\bibfnamefont {K.~N.}\ \bibnamefont {{Yakunin}}},\ }\href {\doibase
  10.1088/2041-8205/807/2/L31} {\bibfield  {journal} {\bibinfo  {journal}
  {\apjl}\ }\textbf {\bibinfo {volume} {807}},\ \bibinfo {eid} {L31} (\bibinfo
  {year} {2015})}\
  \BibitemShut {NoStop}%
\bibitem [{\citenamefont {Horowitz}\ \emph {et~al.}(2017)\citenamefont
  {Horowitz}, \citenamefont {Caballero}, \citenamefont {Lin}, \citenamefont
  {O'Connor},\ and\ \citenamefont {Schwenk}}]{Horowitz.Caballero.ea:2017}%
  \BibitemOpen
  \bibfield  {author} {\bibinfo {author} {\bibfnamefont {C.~J.}\ \bibnamefont
  {Horowitz}}, \bibinfo {author} {\bibfnamefont {O.~L.}\ \bibnamefont
  {Caballero}}, \bibinfo {author} {\bibfnamefont {Z.}~\bibnamefont {Lin}},
  \bibinfo {author} {\bibfnamefont {E.}~\bibnamefont {O'Connor}}, \ and\
  \bibinfo {author} {\bibfnamefont {A.}~\bibnamefont {Schwenk}},\ }\href
  {\doibase 10.1103/PhysRevC.95.025801} {\bibfield  {journal} {\bibinfo
  {journal} {Phys. Rev. C}\ }\textbf {\bibinfo {volume} {95}},\ \bibinfo
  {pages} {025801} (\bibinfo {year} {2017})}\BibitemShut {NoStop}%
\bibitem [{\citenamefont {Lohs}(2015)}]{Lohs:2015}%
  \BibitemOpen
  \bibfield  {author} {\bibinfo {author} {\bibfnamefont {A.}~\bibnamefont
  {Lohs}},\ }\href@noop {} {Ph.D. thesis},\ \bibinfo  {school} {Technische
  Universit{\"a}t Darmstadt} \bibinfo {year} {2015} (unpublished)
  \BibitemShut {NoStop}%
\end{thebibliography}
\end{document}